\documentclass{elsart}
\usepackage{epsfig}

\newcommand{\glen}{G$_{300}$}
\newcommand{\glenpi}{G$^\pi_{300}$}
\newcommand{\UVU}{UV$_{14}$+UVII}

\newcommand{\KB}{G$^\mathrm{K240}_\mathrm{B180}$}
\newcommand{\KM}{G$^\mathrm{K240}_\mathrm{M78}$}
\newcommand{\capt}[1]{\caption[]{#1}}
\newcommand{\gccm}{\mathrm{\,g\,cm}^{-3}}
\newcommand{\sfs}{$^1\mathrm{S}_0$}
\newcommand{\sfp}{$^3\mathrm{P}_2$}
\newcommand{\comments}[1]{\bigskip\parbox[t]{0.9\linewidth}%
{\small{#1}}}
\newcommand{\eqref}[1]{(\ref{#1})}

\begin{document}


\fbox{\parbox[t]{7cm}{\scriptsize Submitted to Nuclear Physics A}}
\bigskip

\begin{frontmatter}

\title{Thermal Evolution of Compact Stars}

\author{Christoph Schaab}
\address{Institut f{\"u}r theoretische Physik,
  Ludwig-Maximilians Universit{\"a}t M{\"u}nchen, Theresienstr. 37,
  D-80333 M{\"u}nchen, Germany \\
  email: schaab@gsm.sue.physik.uni-muenchen.de}
\author{Fridolin Weber}
\address{Institut f{\"u}r theoretische Physik,
  Ludwig-Maximilians Universit{\"a}t M{\"u}nchen, Theresienstr. 37,
D-80333 M{\"u}nchen, Germany}
\author{Manfred K. Weigel}
\address{Sektion Physik der Ludwig-Maximilians Universit{\"a}t 
	M{\"u}nchen, 
Am Coulombwall 1, D-85748 Garching, Germany}
\author{Norman K. Glendenning}
\address{Nuclear Science Division and Institute for Nuclear 
	\& Particle 
Astrophysics \\ Lawrence Berkeley National Laboratory, Berkeley, CA
  94720, U.S.A.}

\begin{abstract}
  A collection of modern, field-theoretical equations of state is
  applied to the investigation of cooling properties of compact 
  stars.
  These comprise neutron stars as well as hypothetical strange matter
  stars, made up of absolutely stable 3-flavor strange quark matter.
  Various uncertainties in the behavior of matter at supernuclear
  densities, e.g., hyperonic degrees of freedom, behavior of coupling
  strengths in matter, pion and meson condensation, superfluidity,
  transition to quark matter, absolute stability of strange quark
  matter, and last but not least the many-body technique itself are
  tested against the body of observed cooling data.
\end{abstract}


\bigskip

\begin{small}
\begin{center}
\end{center}
\end{small}

\end{frontmatter}


\section{Introduction}

A forefront area of research, both experimental and theoretical,
concerns the exploration of the properties of matter under extreme
conditions of temperature and/or density and the determination of the
equation of state (pressure versus density) associated with it. Its
knowledge is of key importance for our understanding of the physics of
the early universe, its evolution to the present day, compact stars,
various astrophysical phenomena, and laboratory physics (for an
overview, see, for example, \cite{Greiner94a}).  On the earth,
relativistic heavy-ion colliders provide the only tool by means of
which such matter can be created and its properties studied.  On the
other hand, however, it is well known that nature has created a large
number of massive stellar objects, i.e., white dwarfs and neutron
stars, which contain matter in one of the densest forms found in the
universe.  Neutron stars are associated with two classes of
astrophysical objects -- pulsars and compact X-ray sources.  Matter in
their cores possess densities ranging from a few times the density of
normal nuclear matter to about an order of magnitude higher, depending
on star mass.  To the present day, about 600 pulsars are known, and
the discovery rate of new ones is rather high. This is accompanied by
an impressive growth rate of the body of observed pulsar data, like
pulsar temperatures determined by the X-ray observatories Einstein,
EXOSAT, and ROSAT \cite{Truemper92a,Truemper93a,Oegelman95a}.

In this paper, we shall apply a broad collection of modern,
field-theoretical equations of state (EOS) to the study of the cooling
behavior of both neutron stars and their strange counterparts --
strange matter stars -- which should exist if 3-flavor strange quark
matter is more stable than confined hadronic matter. This collection
of EOSs was derived under numerous model assumptions about the
behavior of superdense stellar matter.  To mention several are: the
many-body technique used to determine the equation of state; the model
for the nucleon-nucleon interaction; description of electrically
charge neutral neutron star matter in terms of either only neutrons,
neutrons and protons in generalized chemical equilibrium ($\beta$
equilibrium) with electrons and muons, or nucleons, hyperons and more
massive baryon states in $\beta$ equilibrium with leptons; behavior of
the hyperon coupling strengths in matter, inclusion of meson ($\pi$,
$K$) condensation; treatment of the transition of confined hadronic
matter into quark matter; and assumptions about the true ground state
of strongly interacting matter (i.e., absolute stability of strange
quark matter relative to baryon matter).

The paper is organized as follows.  In section \ref{sec:structure} we
introduce the set of equations that govern the cooling behavior of
massive stars. The collection of equations of state for neutron stars
is discussed in section \ref{sec:models}.  The physics of strange
stars and their associated EOS is explained in section
\ref{sec:strangestars}. The phenomenon of superfluidity and the
various neutrino emission processes are outlined in sections
\ref{sec:superfl} and \ref{sec:neutrino}. Our results and conclusions
are presented in sections \ref{sec:res} and \ref{sec:con},
respectively. We summarize the paper in section \ref{sec:summary}.


\section{Structure equations}\label{sec:structure}

The structure of compact, massive stars is determined by Einstein's
field equations,
\begin{equation}
  G^{\mu\nu} = 8\pi T^{\mu\nu} ~,
\label{eq:gmunu}
\end{equation} where $G^{\mu\nu}$ denotes the Einstein tensor and
$T^{\mu\nu}$ is the energy-momentum tensor. For spherically symmetric,
non-rotating stellar configurations the line element is given by
(Schwarzschild metric)
\begin{equation}
  \d s^2 = -\e^{2\phi}\d t^2 + \e^{2\Lambda}\d r^2 + r^2\d\theta^2 +
  r^2 \sin^2\theta\d\varphi^2 \, ,
\label{eq:ds}
\end{equation} where $\phi(r)$ and $\Lambda(r)$ denote the radially
dependent metric functions.

Equations (\ref{eq:gmunu}) and (\ref{eq:ds}) combined with the law of
energy-momentum conservation, $T^{\mu\nu}_{~~;\nu}=0$, lead to
\begin{eqnarray}
  \frac{\d r}{\d N} &= 
	& \frac{1}{4\pi r^2 n \e^\Lambda}~, \label{eq:tov.r} \\
\nonumber \\
  \frac{\d M}{\d N} &= 
	& \frac{\rho}{n} \e^{-\Lambda}~, \label{eq:tov.m} \\
\nonumber \\
  \frac{\d\phi}{\d N} &= 
	& \frac{G\left(\frac{4\pi}{c^2}r^3P+M\right)\e^\Lambda}{4\pi
	c^2r^4n}~, \label{eq:tov.phi} \\
\nonumber \\
  \frac{\d P}{\d N} &= 
	& -(P+c^2\rho)\frac{\d\phi}{\d N}~, \label{eq:tov.p}
\end{eqnarray} which describe the structure of static stars that are in
general-relativistic hydrostatic equilibrium
(Tolman-Oppenheimer-Volkoff equations, cf.
\cite{Misner73,Thorne77}). The metric function $\Lambda$ is determined
by
\begin{equation}
  \e^\Lambda = \left(1-\frac{2GM}{rc^2}\right)^{-\half}~.
\end{equation} 

The baryon number $N$ is introduced as an integration variable, since
this is the only conserved quantity in relativistic field theory.
$M(N)$ is the total mass within a sphere that contains $N$ baryons. Its
surface area is equal to $4\pi r^2(N)$. The quantity $P(N)$ denotes
pressure as a function of baryon number, $\rho(P(N))$ is the total energy
density, and $n(P(N))$ the baryon number density. Finally, we mention
that the functions $\rho(P)$ and $n(P)$ are related to each other by
the equation of state (EOS).

The corresponding boundary conditions are given by
\begin{eqnarray}
  r(N=0) &=& 0\, , \\
  M(N=0) &=& 0\, , \\
  \phi(N=N_\mathrm{m}) &=& 
  \half \ln \left(1-\frac{2GM(N=N_\mathrm{m})}{r(N=N_\mathrm{m})c^2}\right) \,
 , \\
  \rho(P(N=N_\mathrm{m})) &=&\rho_\mathrm{m} \, .
\end{eqnarray} The quantity $N_\mathrm{m}$ is defined by the relation
\begin{equation}
  M(N=N_\mathrm{m}) = M_\mathrm{m}\, ,
\end{equation}
where $M_\mathrm{m}$ denotes a given star mass.

The photosphere of the star is treated separately
from the structure equations (\ref{eq:tov.r})--(\ref{eq:tov.p}) for
two reasons: Firstly, the equation of state of the photosphere depends
much more sensitively on temperature than the core's equation of
state, which can be treated at zero temperature; secondly, because of
a scaling behavior (see, e.g., \cite{Gudmundson83,VanRiper88}) the
structure equations of the photosphere need to be solved only once,
for a properly chosen star mass. The interface between crust and
photosphere is located at the density $\rho_\mathrm{m}$. We shall use
the value given by Van Riper \cite{VanRiper88},
$\rho_\mathrm{m}=10^{10}$~g\,cm$^{-3}$.

Since the central star temperature drops down to less than
$10^9~\mathrm{K}$ within a few minutes after birth \cite{Burrows86},
the effects of finite temperatures on the equation of state can be
neglected to a very good approximation. Consequently, the four
differential equations (\ref{eq:tov.r})--(\ref{eq:tov.p}) do not depend
on time and thus need to be solved only once, at the beginning of the
numerical cooling simulation.

The thermal evolution is described by the partial differential
equations of energy balance \cite{Thorne77},
\begin{eqnarray}
  \frac{\partial(L\e^{2\phi})}{\partial N} &= 
    &-\frac{1}{n}\left(\epsilon_\nu\e^{2\phi}
  +c_\mathrm{v}\frac{\partial(T\e^\phi)}{\partial t}
    \right)\, , \label{eq:ebal}
\end{eqnarray}
and thermal energy transport,
\begin{eqnarray}
  \frac{\partial(T\e^{\phi})}{\partial N} &= 
    &- \frac{(L\e^{2\phi})\e^{-\phi}}{16\pi^2r^4\kappa n} \, . \label{eq:etran}
\end{eqnarray} The microphysical input into Equations (\ref{eq:ebal})
and (\ref{eq:etran}) are the neutrino emissivity per unit volume,
$\epsilon_\nu(P,T)$, heat capacity per unit volume,
$c_\mathrm{v}(P,T)$, and thermal conductivity, $\kappa(P,T)$. The
boundary conditions for (\ref{eq:ebal}) and (\ref{eq:etran}) read
\begin{eqnarray}
  L(N=0) &= &0 \, , \label{eq:bdl}, \\
  T(N=N_\mathrm{m}) &= &T_\mathrm{m}(r_\mathrm{m},L_\mathrm{m},M_\mathrm{m}) 
  \, , \label{eq:bdt}
\end{eqnarray} where
$T_\mathrm{m}(r_\mathrm{m},L_\mathrm{m},M_\mathrm{m})$ is fixed by the
properties of the photosphere at $r=r_\mathrm{m}$
\cite{Gudmundson83,VanRiper88}.  The initial condition for the star's
temperature can be assumed, without loss of generality, to be
$T(r)\equiv 10^{11}$~K, because the star's thermal evolution at times
greater than say a few months does not depend on the exact initial
temperature profile inside the star.  Equations (\ref{eq:ebal}) and
(\ref{eq:etran}) were solved numerically by means of a
Newton-Raphson-like algorithm.  Table \ref{tab:inputs} summarizes the
\begin{table}[tbp]
  \centering \small \capt{Input quantities for the cooling simulations
    performed in this paper \label{tab:inputs}}
\begin{tabular}{ll}
  Parameter & References \\
  \hline \hline
  equation of state: \\
  \quad crust        & \cite{Baym71,Negele73} \\
  \quad core         & see table \ref{tab:eos} \\
  \hline
  Superfluidity      & see table \ref{tab:sf} \\
  \hline
  Heat capacity      & \cite{VanRiper91,Shapiro83} \\
  \hline
  Thermal conductivity: \\
  \quad crust        & \cite{Itoh84a,Itoh83a,Mitake84} \\
  \quad core         & \cite{Gnedin95a,Haensel91} \\
  \hline
  Neutrino Emissivity in the crust: \\
  \quad pair-, photon-, plasma-processes & \cite{Itoh89} \\
  \quad bremsstrahlung                  & \cite{Itoh83b,Pethick94b} \\
  in the core:                          & see table \ref{tab:emis} \\
  \hline
  photosphere (no magnetic field)       & \cite{VanRiper88} \\
\end{tabular}
\end{table}
microphysical input quantities which enter the set of thermal
structure equations, (\ref{eq:tov.r})--(\ref{eq:tov.p}),
(\ref{eq:ebal}) and (\ref{eq:etran}). The input quantities will be
discussed in detail in sections \ref{sec:models}--\ref{sec:neutrino}.


\section{Models for the equation of state}\label{sec:models}

\subsection{General remarks}

The equation of state (i.e., pressure and composition as function of
energy density) of neutron star matter is the basic input quantity
whose knowledge over a wide range of densities, ranging from the
density of iron at the star's surface up to about 15 times the
density of normal nuclear matter reached in the cores of the most
massive star of a sequence, is necessary when solving the thermal
structure equations.  Since one is dealing with highly
isospin-asymmetric, net-strangeness carrying matter in
$\beta$-equilibrium whose properties cannot be explored in laboratory
experiments, one is left with models for the equation of state which
depend on theoretically motivated assumptions and/or speculations
about the behavior of superdense matter. Sources of uncertainty
concern, for instance, a competition between non-relativistic
Schroedinger-based treatments versus relativistic field theoretical
ones; moreover there exists even a considerable uncertainty with
respect to the adequate many-body technique, which is to be introduced
when solving the coupled equations of motion of many-baryon matter.
Other items of uncertainty concern the baryon/meson composition of
neutron star matter. This question has been treated in the literature
at different levels of complexity. The simplest description
approximates neutron star matter by pure neutron matter, which however
is certainly not the ground state of neutron star matter. In fact
neutron matter will quickly decay by means of the weak force into
chemically equilibrated neutron star matter, whose fundamental
constituents--besides neutrons--are protons, hyperons and possibly
more massive baryons that become populated up to the highest densities
reached in the cores of neutron stars.  Finally we mention the
possible transition of confined baryonic matter into quark matter,
which, as demonstrated recently by one of us \cite{Glendenning92b}, may set
in at a density less than twice normal nuclear matter density. Such
densities are easily reached in neutron stars possessing canonical
masses of $1.4M_\odot$.  Last but not least, there is the open problem
concerning $K$-- or $\pi$--meson condensates in neutron stars.

The cross section of a neutron star can be split roughly into three
distinct regimes. The first one is the star's outer crust, which
consists of a lattice of atomic nuclei and a Fermi liquid of
relativistic, degenerate electrons. The outer crust envelopes what is
called the inner crust, which extends from neutron drip density,
$\rho=4.3\times 10^{11}\gccm$, to a transition density of about
$\rho_\mathrm{tr}=1.7\times 10^{14}\gccm$ \cite{Pethick95}.  Beyond
$\rho_\mathrm{tr}$ one enters the star's third regime, that is, its
core where all atomic nuclei have dissolved into their constituents,
protons and neutrons. Furthermore, as outlined just above, due to the
high Fermi pressure the core will contain hyperons, eventually more
massive baryon resonances, and possibly a gas of free up, down and
strange quarks. Finally $\pi$-- and $K$--meson condensates may be
found there too.  The equation of state of the outer and inner
crust has been studied in several investigations and is rather well
known.  We shall adopt the models derived by Baym, Pethick and
Sutherland \cite{Baym71} and Negele and Vautherin \cite{Negele73},
respectively, for it. The models for the equation of state of the
star's core will be discussed in detail in the next section. They fall
into two categories, non-relativistic variational equations of state
and the relativistic field theoretical ones.

%
\subsection{Non-relativistic variational approximation} \label{subsec:nonr}

The non-relativistic models for the equation of state that are
considered in our investigation use modern two-body interactions,
$\mathbf{V}_{ij}$, whose parameters are adjusted to the properties of
the deuteron and the free two-nucleon scattering problem.  These
two-body interactions are supplemented with phenomenological three-body
interactions, $\mathbf{V}_{ijk}$. The additional free parameters are
adjusted to the properties of nuclear matter and the properties of
$^3{\rm H}$ and $^4{\rm He}$ \cite{Wiringa88}. The Hamiltonian's
structure is given by
\begin{equation}
    \mathbf{H} \; = \; \sum_i \left( {{-\,\hbar^2}\over{2\, m}} \right)
    \, \nabla_i^2
    \; + \; \sum_{i<j} \mathbf{V}_{ij} \; + 
    \; \sum_{i<j<k} \mathbf{V}_{ijk} \;\; .
\label{eq:hamil}
\end{equation} The many-body method adopted to solve the many-body
Schroedinger equation is the variational approach
\cite{Wiringa88}, where a variational trial wave-function, $|\Psi_v>$, is
constructed from a symmetrized product of two-body correlation
operators, $F_{ij}$, which act on an unperturbed ground-state, i.e.,
\begin{equation}
      |\Psi_v> \; = \; \left[ \mathbf{S} \; \prod_{i<j} \; 
      \mathbf{F}_{ij} \right] |\Phi>  \;\; .
\label{eq:psiv}
\end{equation} The quantity $|\Phi>$ denotes the antisymmetrized
Fermi-gas wave function.  The correlation operator contains
variational parameters which are varied to minimize the energy per
baryon for a given baryon number density $n$ \cite{Wiringa88},
\begin{equation}
      E_v(n) \; = \; \mathrm{min} \;\;\; \left\{
{ {<\, \Psi_v\, |\, H\, |\, \Psi_v\, >}
               \over{<\, \Psi_v\, |\, \Psi_v\, >} } \right\}~.
\label{eq:evar}
\end{equation} Equations (\ref{eq:psiv}) and (\ref{eq:evar}) were
solved in \cite{Wiringa88} simultaneously in combination with the
equations for electric charge neutrality and $\beta$-equilibrium
($q_B$ and $\mu^B$ denote electric charge and chemical potential of
baryon $B$, respectively; $\mu^\lambda$ is the chemical potential of
leptons, $\lambda=e^-,\; \mu^-$),
\begin{equation}
  \sum_{B=p,n,\ldots} q_B (2J_B+1) \frac{k_{F,B}^3}{6\pi^2}\; -\;
  \sum_{L=e,\mu} \frac{k_{F,\lambda}^3}{3\pi^2} = 0 ~,
\label{eq:elec}
\end{equation}
\begin{equation}
  \mu^B = \mu^n - q_B \; \mu^e~ , \qquad \mu^\mu = \mu^e ~,
\label{eq:chem}
\end{equation} which relate the Fermi momenta, $k_{F,B}$ and
$k_{F,\lambda}$ and chemical potentials of the various baryons and
leptons with each other \cite{Weber91}.

As representative models for the equation of state derived within this
framework, we have selected a model derived by Wiringa, Fiks and
Fabrocini \cite{Wiringa88}, where the two-body interaction UV$_{14}$
supplemented with the three-body interaction UVII, derived by the
Urbana group, serve as an input.  This equation of state takes only
protons and neutrons in equilibrium with respect to the weak force
into account.

%
\subsubsection{Relativistic, field theoretical approximations}

The structure of a lagrangian that accounts for the physics of neutron
star matter is given by \cite{Weber91}
\begin{equation}
  \mathcal{L} = \sum_B \mathcal{L}^0_B + \sum_M \mathcal{L}^0_M +
  \sum_{\lambda} \mathcal{L}^0_{\lambda} + \sum_{B,M}
  \mathcal{L}^{\mbox{int}}_{B,M} + \mathcal{L}^{(\sigma^4)} \; ,
\label{eq:lagr}
\end{equation} where $B$ is summed over those baryons that become
populated up to the highest densities reached in the cores of neutron
stars, i.e., $B=\mathrm{n}$, p, $\Sigma^{\pm,0}$, $\Lambda$,
$\Xi^{0,-}$, $\Delta^{-}$. The quantity $M$ sums the mesons that are
exchanged between them, $M=\sigma$, $\omega$, $\pi$, $\rho$, $\eta$,
$\delta$, and $\phi$. The only contributing leptons are electrons and
muons, hence $\lambda=\mathrm{e}^-$, $\mu^-$.  The first three terms
in (\ref{eq:lagr}) are the lagrangians of free particles. The term
$\mathcal{L}^{\mbox{int}}_{B,M}$ describes the interactions between
the baryons, and $\mathcal{L}^{(\sigma^4)}$ represents
self-interactions of $\sigma$ mesons.

By means of applying the Green's function technique to the equations
of motion that follow from (\ref{eq:lagr}), one obtains the following
coupled system of equations \cite{Weber91,Martin59,Huber94,Huber95}.
The first one is Dyson's equation which determines the two-point
Green's functions, $g_1^B$,
\begin{equation} \label{eq:dyson}
  \left( (g^{0,B}_1)^{-1}(1,2)-\Sigma^B(1,2) \right) g^B_1(1,1') =
  \delta(1,1') \; ,
\end{equation} The second equation determines the effective
baryon-baryon scattering amplitude in matter, $T^{BB'}$,
\begin{eqnarray}
  \langle 12|T^{BB'}|1'2'\rangle &=& \langle 12|v^{BB'}|1'2'-2'1'\rangle 
\nonumber \\
&& + \langle 12|v^{B\bar B}|34 \rangle \Lambda^{\bar B\bar B'}(34,56) 
  \langle 56|T^{\bar B'B'}|1'2' \rangle \; ,
\end{eqnarray} where $\Lambda$ stands for the Brueckner propagator
\cite{Weber91,Huber94,Huber95,Poschenrieder88a,Poschenrieder88b}.

The system of equations is closed by the expression for the self energy of
a baryon in matter, $\Sigma^B$,
\begin{equation} \label{eq:selfenergy}
  \Sigma^B(1,2) = -i \langle 13|T^{BB'}|42 \rangle g_1(4,3)~. 
\end{equation} We use the convention to sum or integrate over
doubly occurring variables. The one-boson-exchange potential, $v$, sums
the contributions arising from the various types of mesons,  
\begin{equation}
  \langle 12|v^{BB'}|34 \rangle =
  \sum_{M=\sigma,\omega,\pi,\rho,...} \; \langle 12|v_M^{BB'}|34 \rangle \; .
\end{equation}

To obtain a relativistic model for the equation of state of neutron
star matter, equations \eqref{eq:dyson} to \eqref{eq:selfenergy} are
to be solved self-consistently in combination with relations
(\ref{eq:elec}) and (\ref{eq:chem}) for electric charge neutrality and
chemical equilibrium.  This has been done for different levels of
approximation.  With increasing level of complexity, these are the
relativistic Hartree (RH), relativistic Hartree-Fock (RHF), and
relativistic Brueckner-Hartree-Fock (RBHF) approximations
\cite{Weber91,Weber91b}.  The latter scheme constitutes the most complex
treatment that is presently manageable. It has been applied
successfully to the composition of neutron star matter only recently.
In contrast to RH and RHF, where the parameters of the theory, are to
be adjusted to the properties of infinite nuclear matter, the RBHF
method makes use of one-boson-exchange potentials whose parameters are
adjusted to the properties of the deuteron and the relativistic
nucleon-nucleon scattering problem. This leaves one with a so-called
``parameter-free'' treatment.

%
\subsection{Comparison of the equations of state}\label{ssec:eos:comparison}

\begin{figure}[tbp] \centering 
\psfig{figure=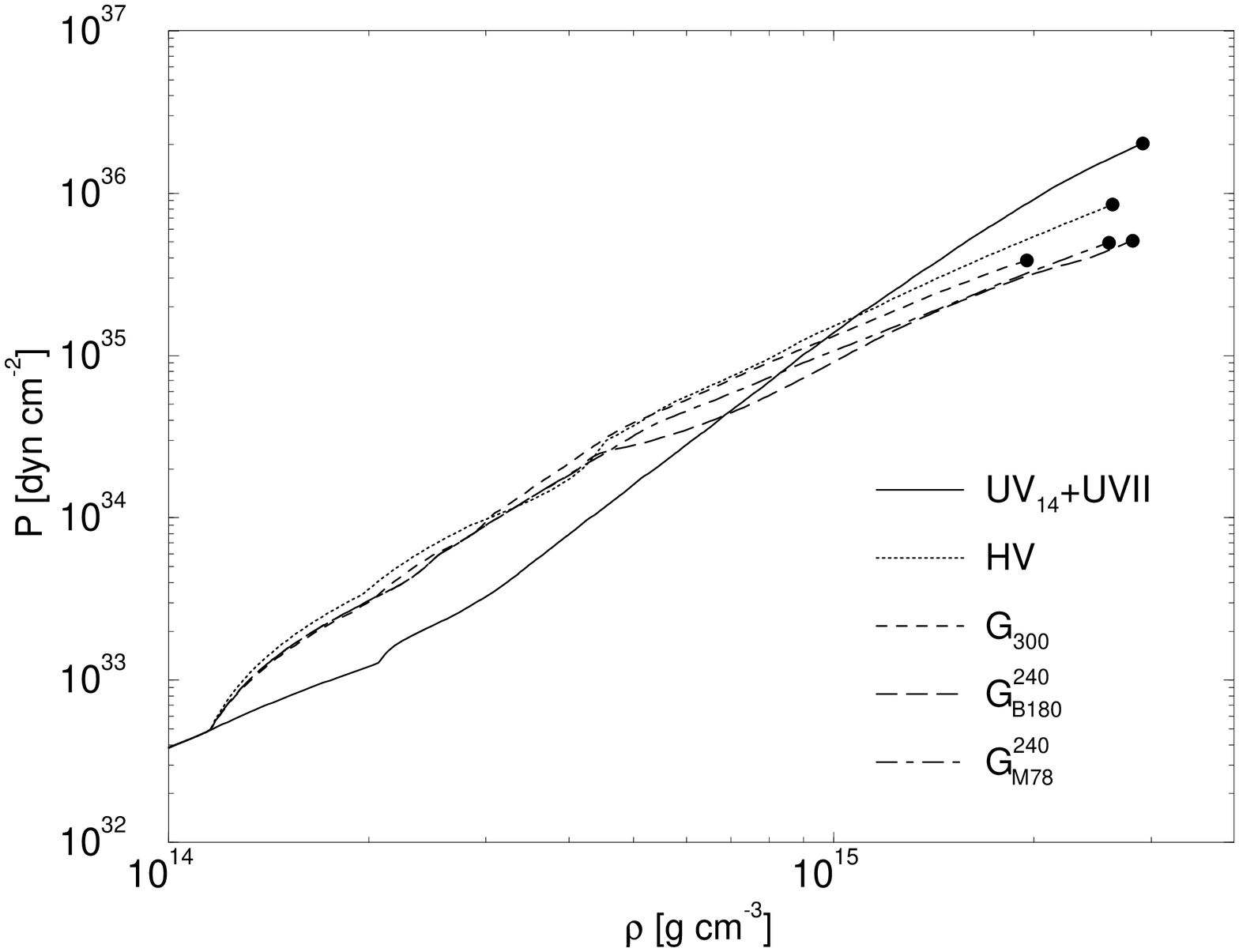,height=10cm,angle=-90}
\capt{Pressure-density relation for several EOSs. The dots mark the
  highest central mass density reached in the most massive neutron
  star of each sequence. Neutron stars with still higher central
  densities are unstable against compressional modes.}
\label{fig:eos}
\end{figure}

An overview of the collection of equations of state used in this paper
is given in table \ref{tab:eos}. This collection serves to study both
\begin{table} 
  \centering \capt{Dynamics and approximation schemes for equations
    of state derived for the cores of neutron stars\label{tab:eos}}
  \smallskip \scriptsize
\begin{tabular}{p{20mm}p{30mm}p{45mm}p{20mm}p{20mm}}
  EOS & Composition & Interaction & Many body approach & Reference \\
  \hline \hline
  UV$_{14}$+UVII 
  & p, n, e$^-$, $\mu^-$ 
  & 2-nuclei potential Urbana V$_{14}$ and 3-nuclei potential Urbana VII 
  & non-relativistic variational method 
  & \cite{Wiringa88} \\
  \hline
  HV
  & p, n, $\Lambda$, $\Sigma^{\pm,0}$, $\Xi^{0,-}$, e$^-$, $\mu^-$
  & Exchange of $\sigma$, $\omega$, $\rho$ mesons
  & RH 
  & \cite{Weber89} \\
  \hline
  RBHF(B)+HFV
  & p, n, $\Lambda$, $\Sigma^{\pm,0}$, $\Xi^{0,-}$, $\Delta^-$, e$^-$, $\mu^-$
  & Exchange of $\sigma$, $\omega$, $\pi$, $\rho$ $\eta$, $\delta$ mesons
  & RBHF 
  & \cite{Huber94,Huber95} \\
  \hline
  \glen
  & p, n, $\Lambda$, $\Sigma^{\pm,0}$, $\Xi^{0,-}$, e$^-$, $\mu^-$
  & Exchange of $\sigma$, $\omega$, $\rho$ mesons
  & RH 
  & \cite{Glendenning89} \\
  \hline
  \glenpi
  & p, n, $\Lambda$, $\Sigma^{\pm,0}$, $\Xi^{0,-}$, e$^-$, $\mu^-$, pion-condensate
  & Exchange of $\sigma$, $\omega$, $\rho$ mesons
  & RH 
  & \cite{Glendenning89} \\
  \hline
  \KB   & p, n, $\Lambda$, $\Sigma^{\pm,0}$, $\Xi^{0,-}$, e$^-$, $\mu^-$, quark matter phase
  & Exchange of $\sigma$, $\omega$, $\rho$ mesons
  & RH, MIT bag model
  & \cite{Glendenning95b} \\
  \hline
  \KM
  & p, n, $\Lambda$, $\Sigma^{\pm,0}$, $\Xi^{0,-}$, e$^-$, $\mu^-$
  & Exchange of $\sigma$, $\omega$, $\rho$ mesons
  & RH
  & \cite{Glendenning95b} \\
  \hline
\end{tabular} \comments{Abbreviations: RH=relativistic Hartree
  approximation, RBHF=relativistic Brueckner-Hartree-Fock
  approximation.}
\end{table}
the influence of competing many-body approximations as well as
different models for the nuclear forces on the cooling behavior of
neutron stars.  Table \ref{tab:eos2} contains the corresponding
\begin{table} 
  \capt{Properties of nuclear matter for the equations of state used
    in this work}\label{tab:eos2} \smallskip \centering
\begin{tabular}{p{4cm}ccccc}
  EOS& $E/A$ [MeV] & $n_0$ [fm$^{-3}$] & $K$ [MeV] & $M^*$ & $a_{\mbox{sy}}$ [MeV] \\
  \hline \hline
  non-relativistic models&        &       &     &      &      \\
  UV$_{14}$+UVII  & -11.5  & 0.175 & 202 & 0.79 & 29.3 \\
  \hline 
  relativistic models   &        &       &     &      &      \\
  HV              & -15.98 & 0.145 & 285 & 0.77 & 36.8 \\
  RBHF(B)         & -15.73 & 0.172 & 249 & 0.73 & 34.3 \\
  \glen           & -16.3  & 0.153 & 300 & 0.78 & 32.5 \\
  \glenpi         & -16.3  & 0.153 & 300 & 0.78 & 32.5 \\
  \KB             & -16.3  & 0.153 & 240 & 0.78 & 32.5 \\
  \KM             & -16.3  & 0.153 & 240 & 0.78 & 32.5 \\
  \hline
\end{tabular} \comments{The entries are: saturation density, $n_0$;
  binding energy per baryon, $E/A$; compression modulus, $K$;
  effective mass in units of the nucleon mass, $M^*=m^*/m$; asymmetry
  energy, $a_{\mbox{sy}}$.}
\end{table}
nuclear matter properties. A graphical illustration of some of the
equations of state is given in fig. \ref{fig:eos}.

With the exception of RBHF(B), the coupling constants of all other
relativistic approximation schemes are chosen such that the properties
of infinite nuclear matter ($n_0\approx (0.15-0.16)$~fm$^{-3}$, $E/A
\approx - 16$~MeV, $m^*\approx 0.78$, $K=200 - 300$ MeV, $a_{\rm
  sy}\approx 32$ MeV) can be reproduced. For RBHF(B), Brockmann's
one-boson-exchange potential (version `B') was used as an input,
which, as pointed out above, leads to a so-called parameter-free treatment.

The ratio of hyperon to nucleon couplings to the meson fields,
\begin{equation}
x_\sigma = g_{H\sigma}/g_{N\sigma}\; , \;\;
x_\omega = g_{H\omega}/g_{N\omega}\; , \;\;
x_\rho   = g_{H\rho}/g_{N\rho}\; , 
\label{eq:ghyp}
\end{equation}
are not defined by the ground-state properties of normal
nuclear matter and so must be chosen according to other considerations
\cite{glenmos91}. The binding energy of 
the $\Lambda$, for example, serves to strictly correlate the values of
$x_\sigma$ and $x_\omega$, but leaves a continuous pairwise ambiguity
which hypernuclear levels and neutron-star masses are able to resolve
\cite{glenmos91}.  In this paper we choose for the hyperon-to-nucleon
scalar coupling $x_\sigma$ a minimum allowable value of $x_\sigma\sim
0.5$ found from the lower bound on the maximum neutron-star mass that
is also consistent with the $\Lambda$ binding in nuclear matter. The
corresponding value of $x_\omega$ amounts $\sim 0.52$.

The possible transition of confined hadronic matter into quark matter
at high densities is contained in the equation of state labeled
\KB~ \cite{Glendenning95b}. The transition
was determined for a bag constant of $B^{1/4}=180$ MeV which places
the energy per baryon of strange quark matter at 1100 MeV, well above
the energy per nucleon in infinite nuclear matter as well as the most
stable nucleus, $^{56}{\rm Fe}$ ($E/A\approx$ 930 MeV).  (The
possibility of absolutely stable strange quark matter will be
considered in sect.  \ref{sec:strangestars}.)  Most interestingly,
this model predicts a transition to quark matter at a density as low
as $1.6\,n_0$. The pure quark-matter phase is reached at a density of
about $10 \, n_0$, which is close to the central density of the
heaviest and thus most compact star constructed from such an equation
of state.  We stress that these density thresholds are rather
different from those obtained by numerous other authors, who performed
work on this topic prior to Glendenning's paper.  The reason for this
lies in the seemingly innocuous assumption of treating neutron star
matter as a simple body, that is, as matter that is characterized by
only one conserved charge.  This is not true for neutron star matter
which possesses two conserved charges, baryonic as well as electric.
Correspondingly, there are two chemical potentials rather than one,
and the transition of baryon matter to quark matter is to be
determined in the three-dimensional space spanned by pressure and the
chemical potentials of the electrons and neutrons.  The only existing
investigations which accounts for this properly are those of
Glendenning \cite{Glendenning92b,Glendenning95b,Glendenning91a}.

\begin{figure}[tbp] \centering 
\psfig{figure=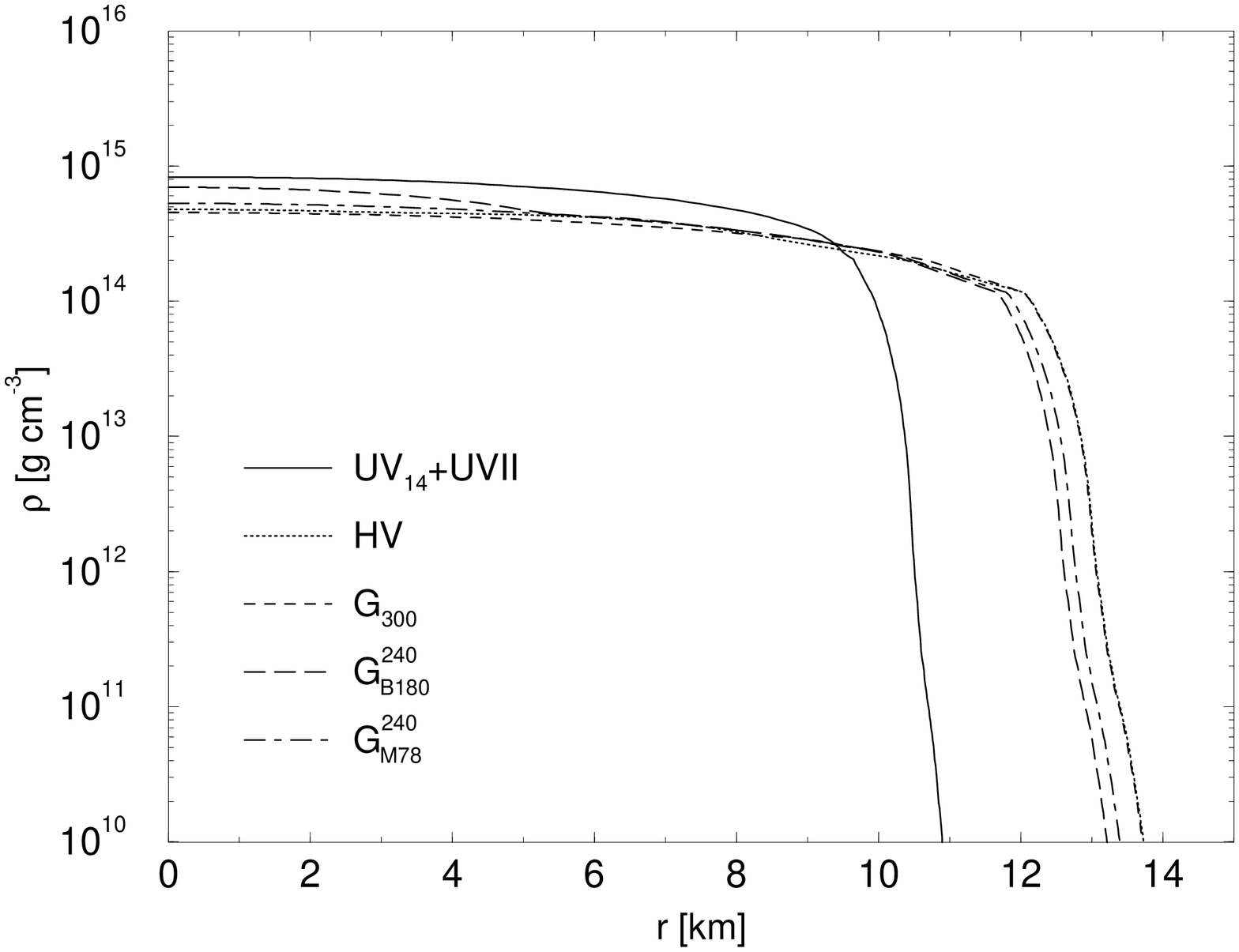,height=10cm,angle=-90}
\capt{Density profiles of neutron star with mass 1.0~$M_{\odot}$
  constructed for a sample of different EOSs.}
\label{fig:rho:r:0.5}
\end{figure}

\begin{figure}[tbp] \centering 
\psfig{figure=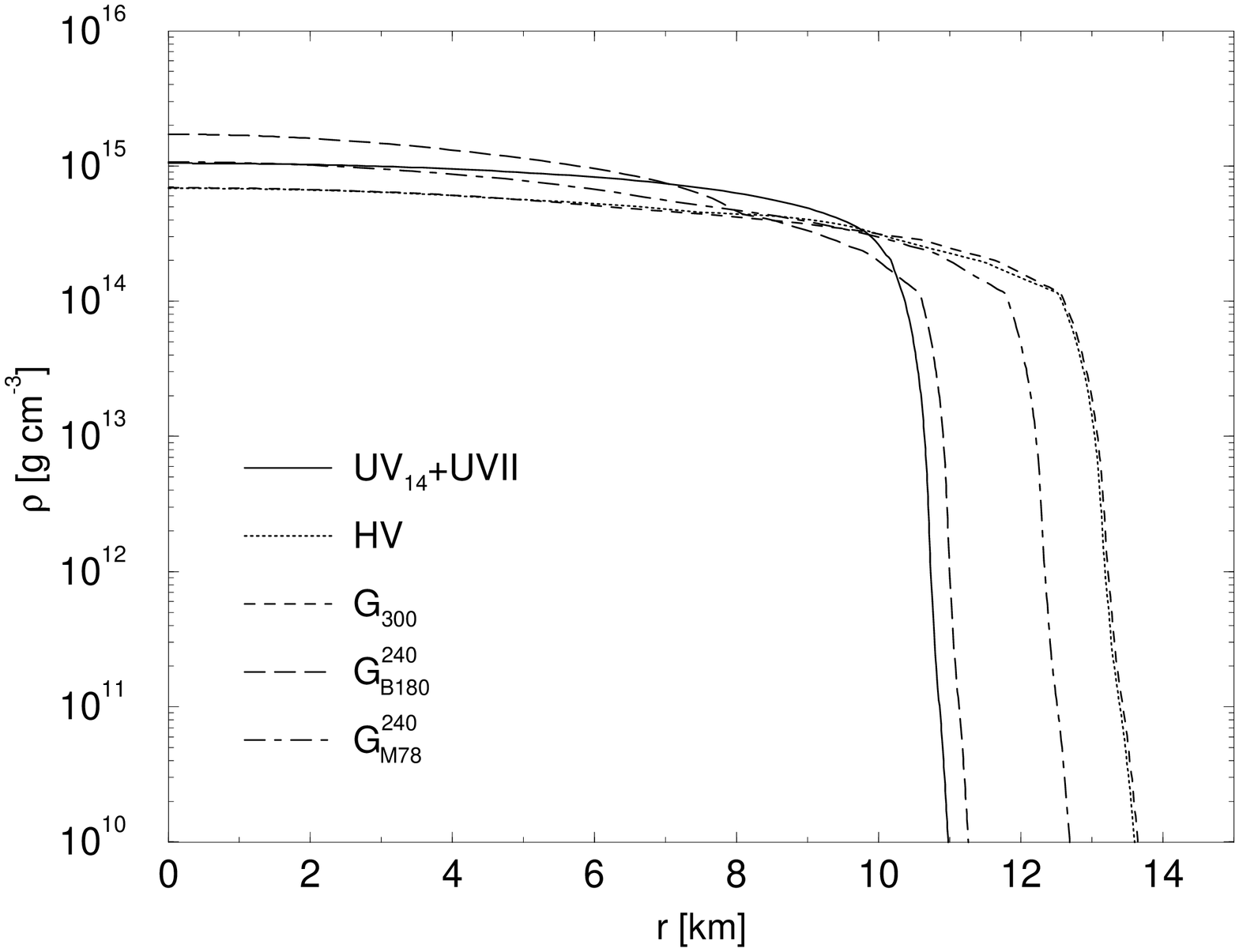,height=10cm,angle=-90}
\capt{Same as fig. \protect{\ref{fig:rho:r:0.5}} but for a neutron
  star with mass 1.4~$M_{\odot}$.}
\label{fig:rho:r:1.4}
\end{figure}

\begin{figure}[tbp] \centering 
\psfig{figure=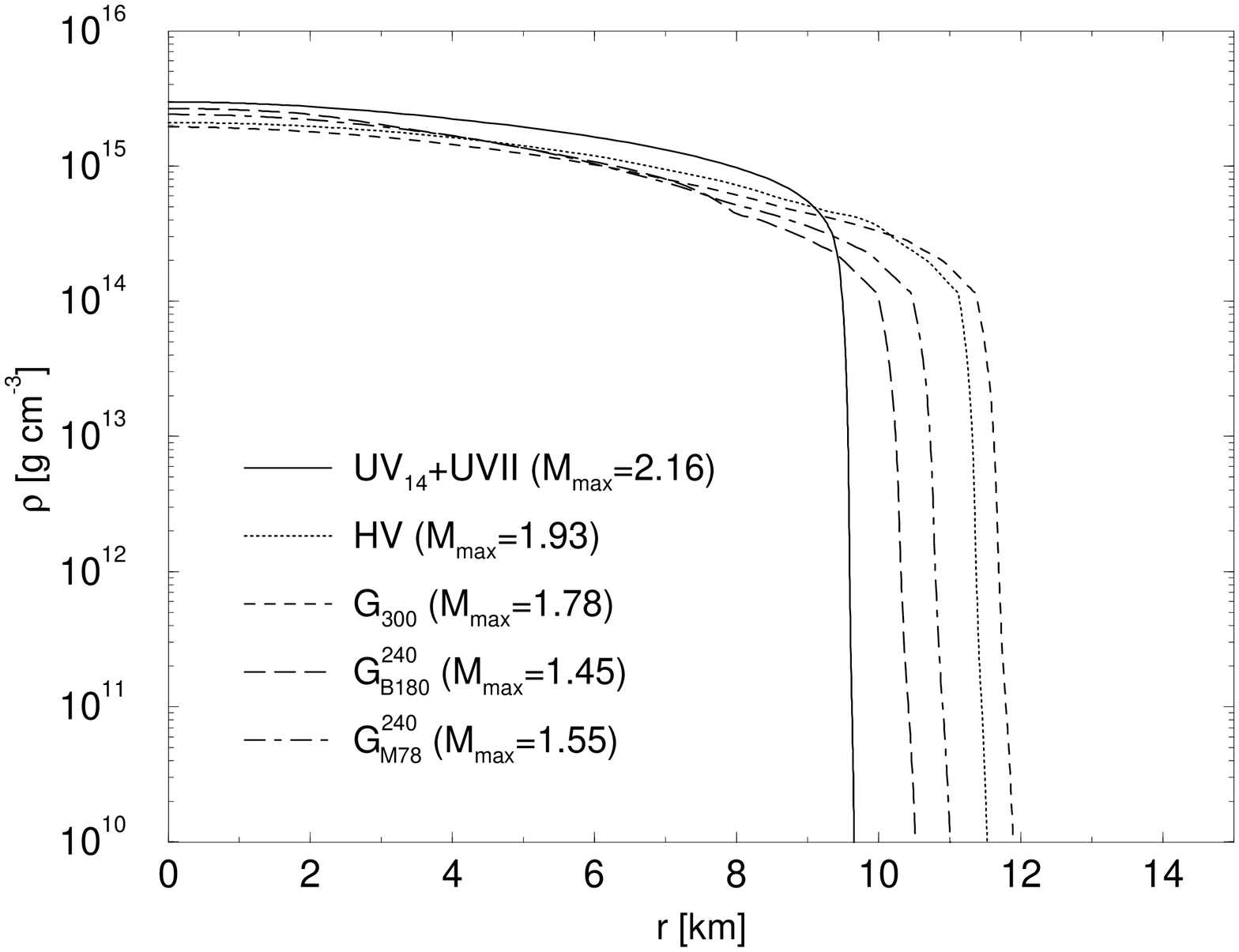,height=10cm,angle=-90}
\capt{Same as fig. \protect{\ref{fig:rho:r:0.5}} but for the heaviest
  neutron star of each sequence.}
\label{fig:rho:r:1.8}
\end{figure}

The stiffness of the equation of state depends strongly on the
internal degrees of freedom. Generally, the more degrees of freedom
accessible to a system, the softer the equation of state. A softer
equation of state, in turn, leads to neutron stars that are less
massive but more dense than those obtained for stiffer equations of
state.  This is illustrated in figs. \ref{fig:rho:r:0.5} to
\ref{fig:rho:r:1.8}, where the density profiles of neutron stars 
constructed for different equations of state are shown. The stiffness
of these equations of state can be inferred from fig. \ref{fig:eos}.

\section{Strange Stars} \label{sec:strangestars}

The hypothesis that strange quark matter may be the absolute ground
state of the strong interaction (not $^{56}{\rm Fe}$) has been raised
independently by Bodmer \cite{Bodmer71} and Witten \cite{Witten84}.
If the hypothesis is true, then \emph{separate} classes of compact
stars could exist, which range from dense strange stars to strange
dwarfs to strange planets
\cite{Weber94,Glendenning94a,Glendenning95c}.  They form distinct and
disconnected branches of compact stars and are not part of the
continuum of equilibrium configurations that include ordinary white
dwarfs and neutron stars. In principle both strange and neutron stars
could exist. However if strange stars exist, the galaxy is likely to
be contaminated by strange quark nuggets which would convert all
neutron stars that they come into contact with to strange stars
\cite{Glendenning90,Madsen91,Caldwell91}.  This would imply that the
objects known to astronomers as pulsars are probably rotating strange
matter stars and not neutron matter stars, as is usually assumed.
Unfortunately, lattice QCD calculations will not be accurate enough in
the foreseeable future to give a definitive prediction as to the
absolute stability of strange matter so that even to the present day
there is not sound scientific basis on which one can either confirm or
reject the hypothesis, so that it remains a serious possibility of
fundamental significance for compact stars plus various other physical
phenomena \cite{Aarhus91}. Below we will explore the cooling behavior
of strange stars and compare it with the one of their non-strange
counterparts, i.e., ordinary neutron stars.  This enables one to test
the possible existence of strange stars and thus draw definitive
conclusions about the true ground-state of strongly interacting
matter.

\begin{figure}[tbp] \centering 
\psfig{figure=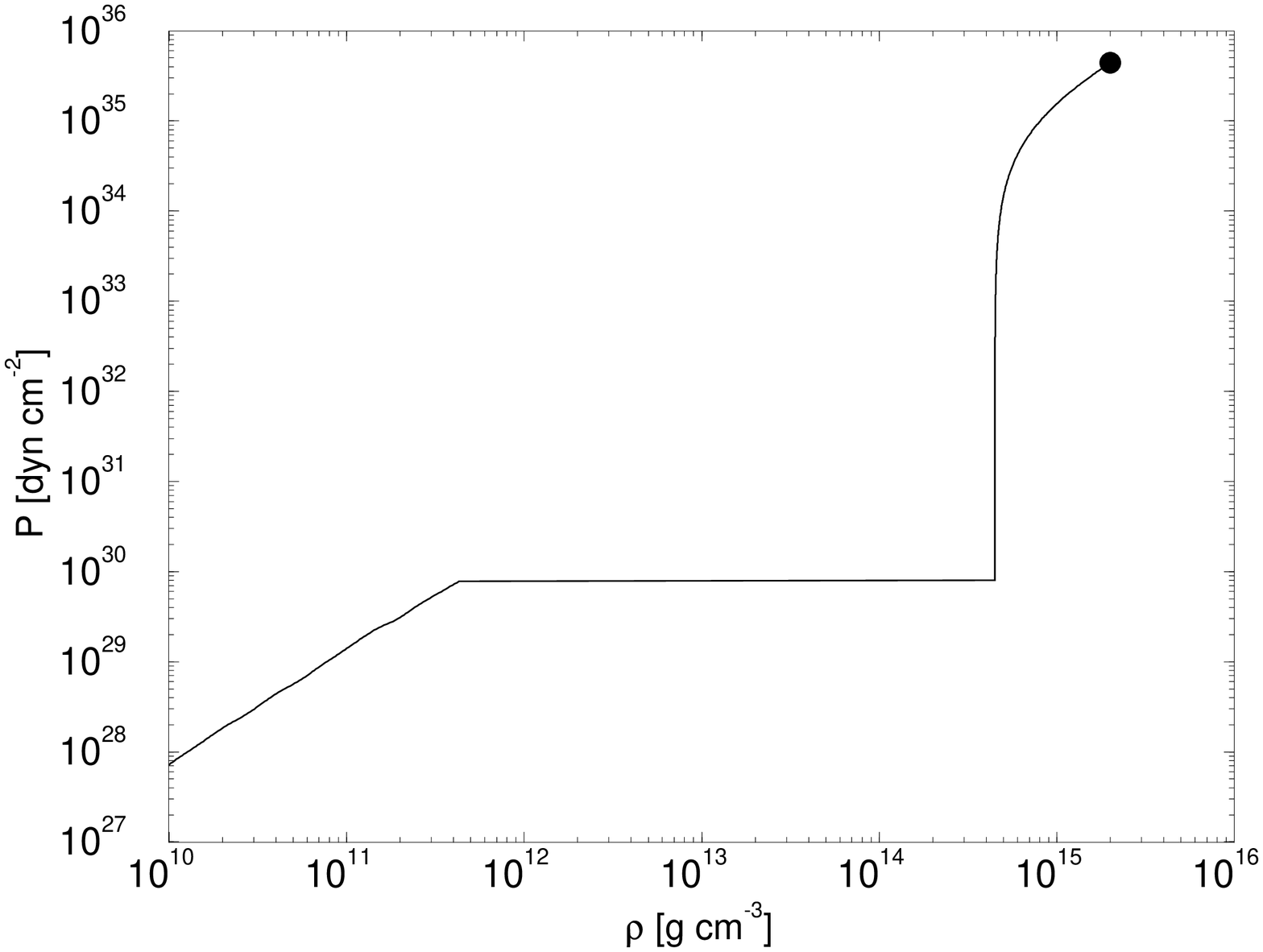,height=10cm,angle=-90}
\capt{Pressure-density relation for strange stars with crust. The bag
  constant is $B^{1/4}=145$~MeV, the mass of the strange quark is
  $m_s=150$~MeV. The solid dot marks the central density reached in
  the heaviest stable strange star. Stars with still higher central
  densities are unstable against compressional modes.}
\label{fig:sm:eos1}
\end{figure}

As pointed out by Alcock, Farhi, and Olinto \cite{Alcock86}, a strange
star can carry a solid nuclear crust whose density at its base is
strictly limited by neutron drip.  This is made possible by the
displacement of electron at the surface of strange matter, which leads
to a strong electric dipole layer there. It is sufficiently strong to
hold open a gap between ordinary atomic (crust) matter and the
quark-matter surface, which prevents a conversion of ordinary atomic
matter into the assumed lower-lying ground state of strange matter.
Obviously, free neutrons, being electrically charge neutral, cannot
exist in the crust, because they do not feel the Coulomb barrier and
thus would gravitate toward the strange-quark matter core, where they
are converted by hypothesis into strange matter.  Consequently, the
density at the base of the crust (inner crust density) will always be
smaller than neutron drip density.  The situation is illustrated in
fig. \ref{fig:sm:eos1} where the equation of state of a strange star
with crust is shown.

The strange-star models presented in the following are constructed for
an equation of state of strange matter derived by Farhi and Jaffe
\cite{Farhi84}. The value chosen for the bag constant and strange
quark mass are $B^{1/4}=145$~MeV and $m_{\rm s}=150$~MeV,
respectively. These choices place the energy per baryon of strange
quark matter at 850 MeV, which corresponds to strongly self-bound
strange matter.

%
\section{Superfluidity}\label{sec:superfl}

The effective nucleon-nucleon interaction contains a rather strong
repulsive, short-range part and a weaker, attractive long-range part.
Hence, depending on the average distance of the particles, nuclear
matter may become superfluid (see \cite{Takatsuka93b} for a recent 
review). Neutrons, for instance, are believed to form \sfs~superfluid
pairs in the density range between neutron drip and and the saturation
density of normal nuclear matter (see below).  At supernuclear
densities, i.e., in the core region of neutron stars, the repulsive
interaction becomes dominant and the \sfs~ superfluidity gap closes.
However, the tensor and spin-orbit interaction become attractive at
such densities and the formation of \sfp -neutron pairs becomes
possible. Finally, the proton concentration in the core region may
reach a partial density simlar to that of the pair-condensed neutron
gas at subnuclear densities, enabling the protons to form superfluid
\sfs~pairs.  Since the protons are charged, the proton superfluid is a
superconductor.  The different superfluid regimes can be summarized as
follows \cite{Ainsworth89a,Amundsen85,Wambach91a}:
\begin{enumerate}
\item $7\times 10^{11}\gccm$ to $2\times 10^{14}\gccm$: ~~neutrons in
  \sfs -pair state,
\item $2\times 10^{14}\gccm$ to $4\times 10^{14}\gccm$: ~~protons in
  \sfs -pair state, 
\item $2\times 10^{14}\gccm$ to $5\times 10^{16}\gccm$: ~~neutrons in
  \sfp -pair state.
\end{enumerate} One has to keep in mind that these values depend on the
composition of matter.  The above values are computed for the equation
of state labeled HV.  A further possibility not mentioned so far are
$^3\mathrm{D}_2$ neutron--proton pairs.  Finally, it is possible that
hyperons and quarks form superfluid states too
\cite{Bailin84,Page95}. Until now, however, no detailed calculations
on this topic exist and one is thus left with parameter studies
(cf. section \ref{ssec:ecquark}).

\begin{figure}[tbp] \centering 
\psfig{figure=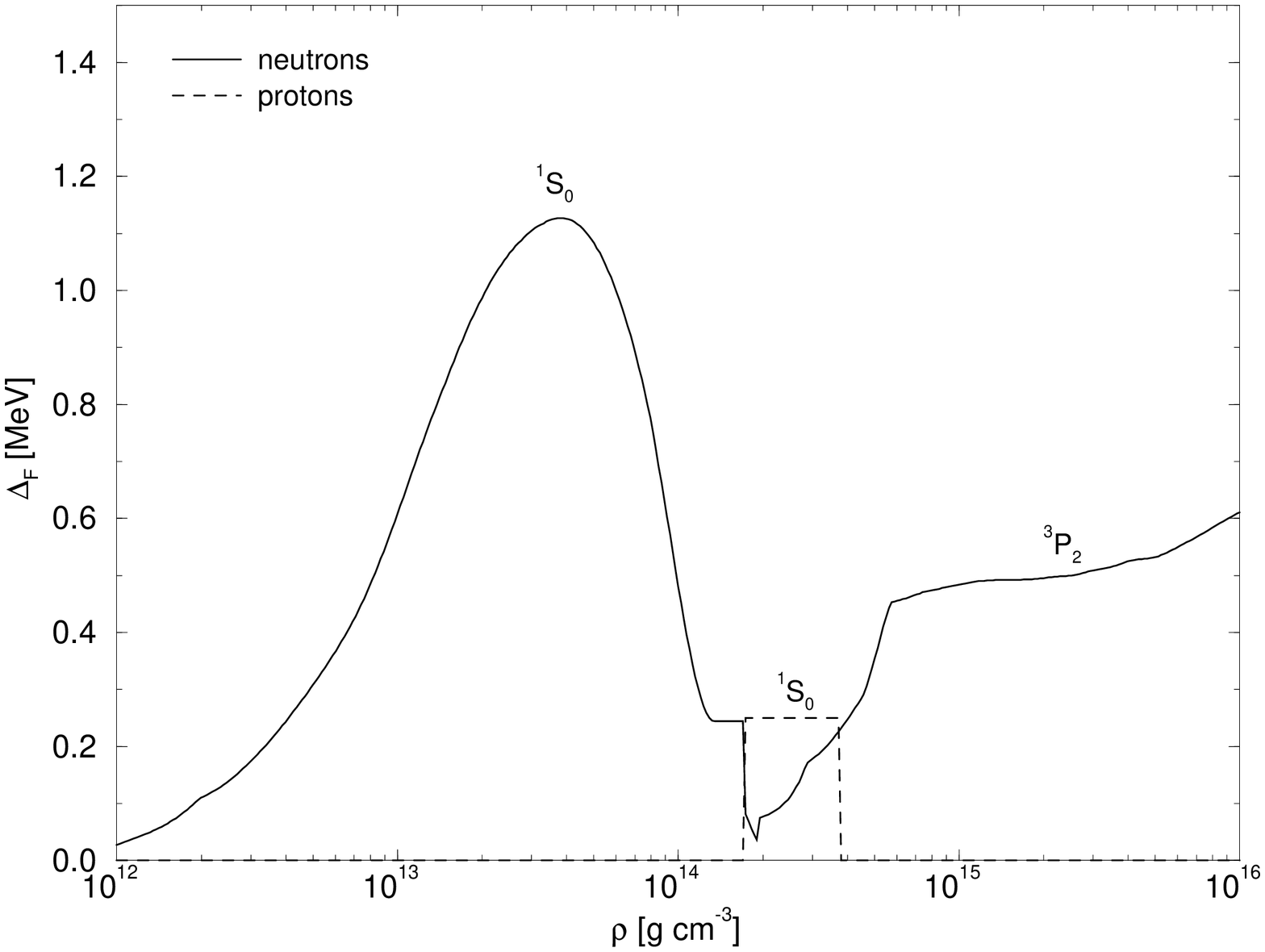,height=10cm,angle=-90}
\capt{Energy gap versus mass density computed for the HV EOS.}
\label{fig:sf}
\end{figure}

If protons and/or neutrons become superfluid, the neutrino emissivity,
thermal conductivity, and heat capacity are reduced by an exponential
factor of the form\footnote{For the \sfp~state this is rigorously true
  only for isotropic matter \cite{Page95,Anderson61,Muzikar80}.}
$\exp(-\Delta_\mathrm{sf} /k_\mathrm{B}T)$, where $\Delta_\mathrm{sf}$
is the gap energy \cite{Gnedin95a,Maxwell79}.  Unfortunately, the gap
energies are not very well known because of uncertainties in the
effective baryon masses and polarization effects \cite{Chen93}. We
will take the gap energies for protons from \cite{Wambach91a}, and
those for the \sfs~ and \sfp~ states of neutrons from
\cite{Ainsworth89a} and \cite{Amundsen85}, respectively. An overview
of the gap energies and critical temperatures is put together in table
\ref{tab:sf}. As an example, we show in fig. \ref{fig:sf} the
\begin{table} 
  \capt{Gap energies and critical temperatures of superfluid states in
    neutron star matter}
\label{tab:sf}
\bigskip \centering
\begin{tabular}{cccc}
                              & Neutron \sfs & Neutron \sfp & Proton \sfs \\
  \hline \hline
  Reference                   & \cite{Ainsworth89a} & \cite{Amundsen85} & \cite{Wambach91a} \\
  \hline
  $\Delta_\mathrm{sf}^\mathrm{max}$ [MeV] & 1.13          & 0.62          & 0.25 \\
  $T^\mathrm{max}_c$ [$10^9$~K] & 7.4       & 0.8           & 1.6 \\
  \hline
  from $\rho$ [$\gccm$]       & $7\times 10^{11}$   & $2\times 10^{14}$   & $2\times 10^{14}$ \\
  to $\rho$ [$\gccm$]         & $2\times 10^{14}$   & $5\times 10^{16}$   & $4\times 10^{14}$ \\
  \hline
\end{tabular} \comments{The density ranges are calculated for the HV
equation of state. $\Delta_\mathrm{sf}^\mathrm{max}=1.13$~MeV is
obtained for the \sfs~superfluid gap as estimated by Ainsworth et al.
Using the gaps estimated in the more recent paper by Wambach et al.
\cite{Ainsworth89a} leads to $\Delta_\mathrm{sf}^\mathrm{max}=1.0$~MeV,
which influences the cooling of neutron stars only insignificantly.}
\end{table}
density dependence of $\Delta_\mathrm{sf}$ for one of our model
equations of state. A qualitatively similar behavior is obtained for
the other models.

\section{Neutrino emissivities} \label{sec:neutrino}

\begin{figure}[tbp] \centering 
\psfig{figure=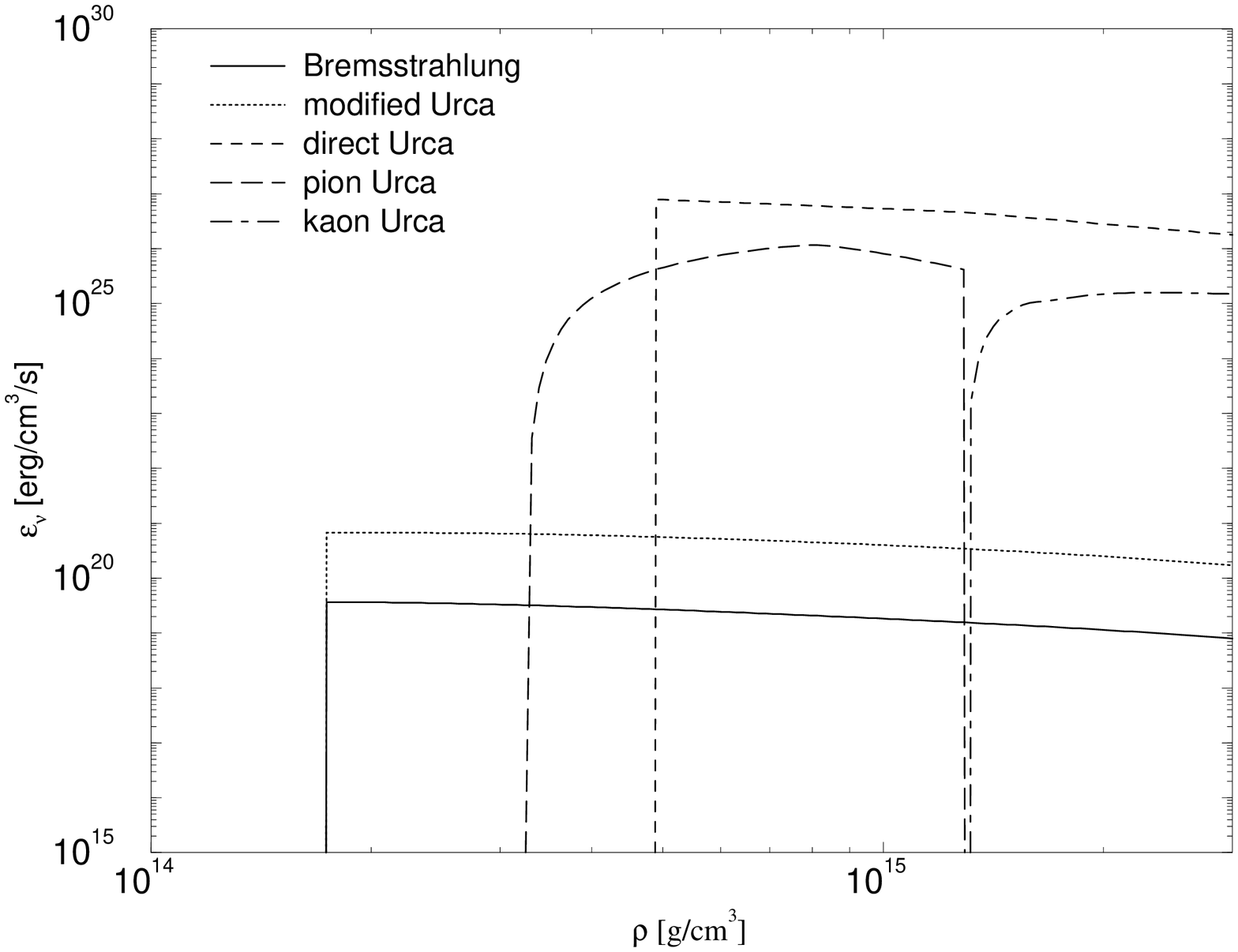,height=10cm,angle=-90}
\capt{Neutrino emissivities of different neutrino-emitting processes as a
  function of energy density for $T=10^9$~K. The particle composition
  is taken from the \glenpi ~EOS.}
\label{fig:emis}
\end{figure}

The neutrino emission processes can be divided into slow and fast
ones. The main feature, which distinguishes fast from slow processes,
is the existence of a certain threshold density for the fast ones,
below which they cannot take place. An overview of all processes which
we shall take into account is given in table \ref{tab:emis}. The
\begin{table} 
  \capt{Neutrino processes studied in this work, their associated
    neutrino emissivities, $\epsilon_\nu$, and the critical threshold
    densities, $n_\mathrm{crit}$, below which these processes are forbidden}
\label{tab:emis}
\smallskip \centering
\begin{tabular}{cccc}
  Process & $\epsilon_\nu$ [erg~s$^{-1}$] & $n_\mathrm{crit}$ [fm$^{-3}$] & Reference \\
  \hline \hline
  Nucleon bremsstrahlung      & $10^{19}\times T_9^8$ &             
        & \cite{Maxwell79,Friman79} \\
  modified Urca       & $10^{21}\times T_9^8$ &             
        & \cite{Friman79}           \\
  direct Urca         & $10^{27}\times T_9^6$ & 0.25 -- 0.5 
        & \cite{Lattimer91}         \\
  Pion-condensate     & $10^{26}\times T_9^6$ & 0.21        
        & \cite{Maxwell77a}            \\
  Kaon-condensate     & $10^{25}\times T_9^6$ & 0.71        
        & \cite{Thorsson94a,Thorsson95a} \\
  Quark $\beta$-decay & $10^{24}\times T_9^6$ & 0.29        
        & \cite{Duncan83}           \\
  \hline
\end{tabular} \comments{The first two processes are slow ones. They
  are possible at all densities, for which bulk nuclear matter exisits. The
  critical density for the direct Urca process depends on the equation
  of state, which leads to a range of threshold densities.}
\end{table}
associated neutrino emissivities are shown in fig. \ref{fig:emis}.

\subsection{Neutrino pair Bremsstrahlung}

The most important neutrino process in the crust is
neutrino-antineutrino pair production by electron bremsstrahlung.  In
early estimates it was found that the contribution of this process to
cooling is competitive with standard cooling via the modified Urca
process.  However, recent investigations have shown that the mass of
the ionic crust is reduced by a factor of two to three if one
considers the recently established value $\rho_\mathrm{tr}\approx 1.7 \times
10^{14}\gccm$ for the transition density between crust and core
\cite{Pethick95}, instead of the older value $\rho_\mathrm{tr}\approx 2.8
\times 10^{14}\gccm$ \cite{Baym71}.  Additionally, considerations by
Pethick and Thorsson \cite{Pethick94b} have pointed out that the
standard perturbative treatment of determining $\epsilon_\nu$ is not
sufficiently accurate below a temperature of $10^{10}$~K.  Furthermore
taking the electronic band structure into account can reduce the
energy emission rates exponentially for temperatures below
$10^{10}$~K. We include these corrections to the emissivities derived 
by Itoh et al. in \cite{Itoh84c,Itoh84d}. Additionally to the above
process, pair, photon and plasmon processes may also be important in
the low density regime. Their effects have been estimated too by Itoh
et al. in \cite{Itoh89}, and will be used in our calculations.

Neutrino-antineutrino pair bremsstrahlung production can also occur in
the core according to
\begin{eqnarray}
  \mathrm{n} + \mathrm{n} &\rightarrow 
    &\mathrm{n} + \mathrm{n} + \nu + \bar{\nu} \; , 
\label{eq:br1} \\
  \mathrm{n} + \mathrm{p} &\rightarrow 
    &\mathrm{n} + \mathrm{p} + \nu + \bar{\nu} \; , 
\label{eq:br2} \\
  \mathrm{e}^- + \mathrm{p} &\rightarrow 
    &\mathrm{e}^- + \mathrm{p} + \nu + \bar{\nu} \; .
\label{eq:br3}
\end{eqnarray} We note that the emissivity of the last bremsstrahlung
process, Equation (\ref{eq:br3}), becomes very important if the
neutrons become superfluid. In this case this process dominates 
all other processes. For the processes (\ref{eq:br1}) to
(\ref{eq:br3}) we use the emissivities as calculated by Friman and
Maxwell \cite{Maxwell79,Friman79}.

%
\subsection{Modified Urca process}

The beta decay and electron capture processes,
\begin{equation} \label{eq:dirUrca}
  \mathrm{n} \rightarrow \mathrm{p} + \mathrm{e}^- + \bar{\nu}_\mathrm{e}
  \qquad \mbox{and} \qquad
  \mathrm{p} + \mathrm{e}^- \rightarrow \mathrm{n}  + \nu_\mathrm{e} \; ,
\end{equation} also known as the direct Urca processes, were not
included in cooling calculations
\cite{Tsuruta66,Tsuruta79a,Glen80,Richardson82,Nomoto87} until
recently, since energy and momentum conservation can only be fulfilled
if the proton fraction exceeds a critical value (cf.
section \ref{subsec:dirUrca}). Most interestingly,
many of the older, non-relativistic models for the equations of state
of neutron star matter indeed do not lead to sufficiently high enough proton
fractions. Consequently, the
so-called modified Urca process, which is characterized by the
occurrence of a bystander particle which is necessary to conserve both
energy and momentum of the baryons that are being scattered, i.e.,
\begin{equation}
  \mathrm{n} + \mathrm{n} \rightarrow 
        \mathrm{n} + \mathrm{p} + \mathrm{e}^- + \bar{\nu}_\mathrm{e} 
  \qquad \mbox{and} \qquad
  \mathrm{n} + \mathrm{p} + \mathrm{e}^- \rightarrow 
        \mathrm{n} + \mathrm{n} + \nu_\mathrm{e}~.
\label{eq:modurc}
\end{equation} were considered to be the dominant cooling mechanism of
neutron stars.  The emission rates associated with processes
(\ref{eq:modurc}) were calculated by Friman and Maxwell
\cite{Friman79}. There the exchanged $\pi$ meson is treated as a free
particle. Such an approximation neglects the influence of the medium
on the propagating pion.  Furthermore, the exchanged pions undergo
weak transitions (so-called exchange current effect). An estimate of
the influence of such effects, performed by Voskresenski{\u i} and
Senatorov \cite{Voskresenskii86,Voskresenskii87a}, shows that the
neutrino emissivities may increase by up to several orders of
magnitude. The implications of these effects of cooling will be
studied in a different investigation \cite{Schaab95b}.

%
\subsection{Direct Urca process} \label{subsec:dirUrca}

\begin{figure}[tbp] \centering 
\psfig{figure=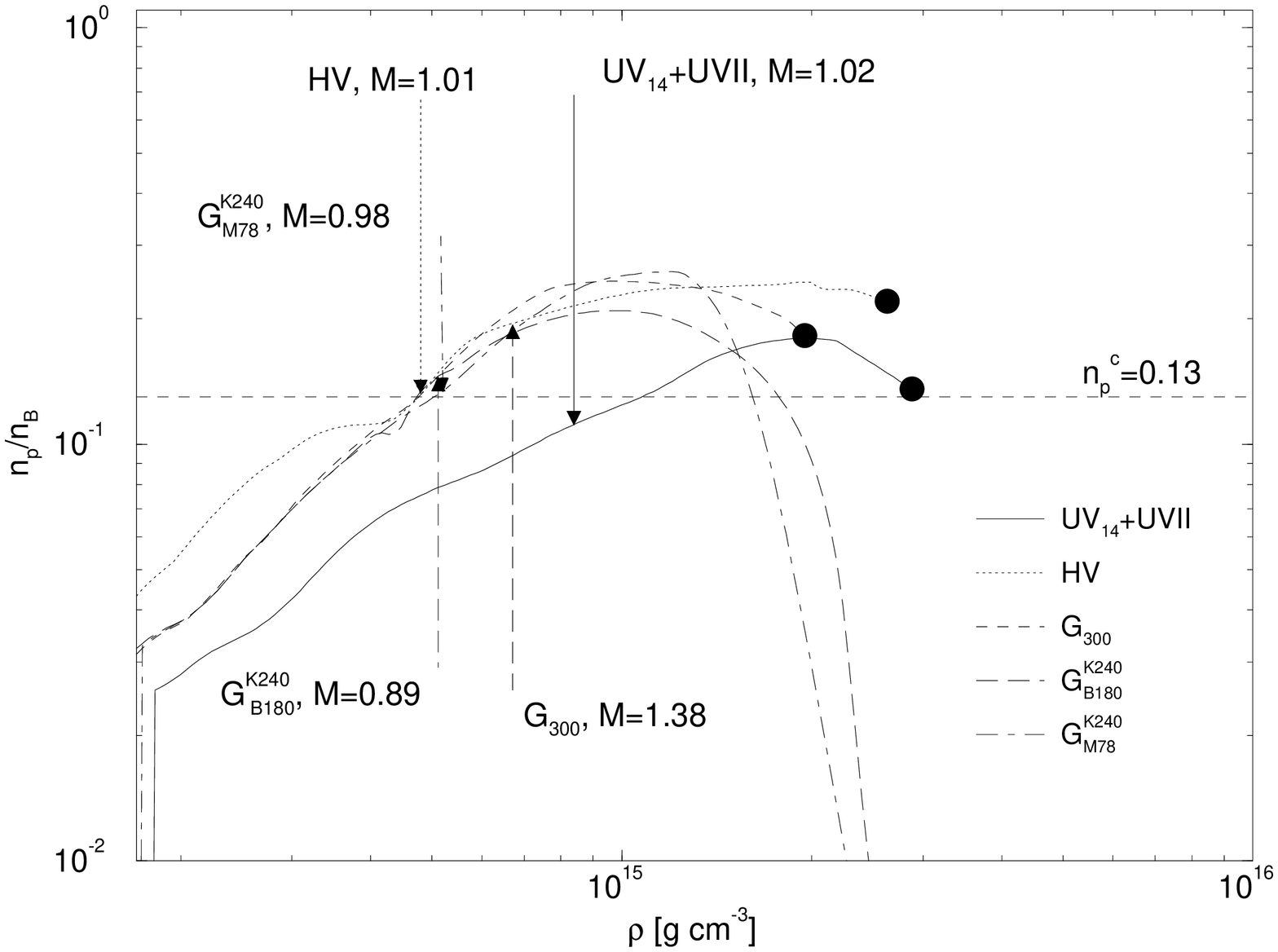,height=10cm,angle=-90}
\capt{Proton fraction in neutron star matter as a function of energy
  density for different equations of state. The approximate critical
  proton fraction $n_\mathrm{p}/n_\mathrm{B}=0.13$ for the direct Urca
  process is marked by the horizontal line. The exact critical energy
  densities together with the corresponding star (masses in units of
  $M_\odot$) are marked by arrows.} \label{fig:protonfraction}
\end{figure}

As already mentioned, the direct Urca process \eqref{eq:dirUrca} is
only possible if the proton fraction exceeds a certain critical value
\cite{Lattimer91,Boguta81a}. If the star is assumed to be made up of
only neutrons, protons and electrons, the critical proton fraction
amounts about $11\%$. In one includes hyperons and muons too, the
critical value increases slightly to $13\%$. In contrast to older,
non-relativistic equations of state which predict proton fractions
that lie below the critical values, modern non-relativistic as well as
relativistic equations of state contain sufficiently large enough
proton fractions such that the direct Urca process becomes possible.
The proton fraction depends crucially on the asymmetry energy which
unfortunately is not known at higher densities.  From our study we
find that the proton fraction is generally larger for the relativistic
equations of state than for the non-relativistic ones.  This trend can
be seen in fig. \ref{fig:protonfraction}, where the proton fractions
are shown for several equations of state of our collection. The
critical energy densities beyond which the direct Urca becomes
possible, are indicated.  Because of the different
asymmetry behaviors of these equations of state, the critical
densities vary considerably from one equation of state to another.
This, in turn, links the onset of the direct Urca process to the
condition of a sufficiently large enough star mass.  For the
emissivities of the direct Urca process we shall use those calculated
by Lattimer et al. \cite{Lattimer91}.  The emissivity of the direct
process exceeds the one of the modified process by up to ten orders of
magnitude, at low temperatures of $T\approx 10^7$~K.

So far we have considered the direct Urca process in neutron star matter
made up of protons and neutrons only. Strange baryons and $\Delta$
isobars, however, which may exist in such matter rather abundantly
too, also produce neutrinos via direct Urca processes of the following
type,
\begin{equation}
  \Sigma^- \rightarrow \Lambda + \mathrm{e}^- + \bar{\nu}_\mathrm{e}
    \qquad \mbox{and} \qquad
  \Lambda + \mathrm{e}^- \rightarrow \Sigma^- + \nu_\mathrm{e}~.
\label{eq:durca2}
\end{equation} Such processes have been studied first by Prakash et
al. \cite{Prakash92}, and the implications for cooling of neutron
stars were explored in \cite{Haensel94:a}.  We have included the above
processes in cooling simulations elsewhere \cite{Schaab95c}. It was
found that the cooling behavior due to direct Urca among the nucleons
only dominates over the energy loss associated with processes the
\eqref{eq:durca2}.  Or, in quantitative terms, the cooling curves
obtained by restriction to the direct Urca process among
non-superfluid nucleons (section
\ref{ssec:edurca}) remain practically unchanged.

%
\subsection{Quasi-particle processes in boson-condensed matter}

In the charged pion ($\pi^\mathrm{c}$-) condensed phase, the
quasi-particle Urca processes
\begin{equation}
  \eta(\pi) \rightarrow \eta(\pi) + \mathrm{e}^- + \bar{\nu}_\mathrm{e}
        \qquad \mbox{and} \qquad
  \eta(\pi) + \mathrm{e}^- \rightarrow \eta(\pi) + \nu_\mathrm{e}
\end{equation} are possible. The quasi-particle state $\eta$ is
given by the superposition of proton and neutron states,
\begin{equation}
  |\eta_\pm\rangle = \cos\theta |\mathrm{n}\rangle 
        \mp i \sin\theta |\mathrm{p}\rangle \; ,
      \end{equation} with the mixing angle $\theta$. The emissivity
for this process was calculated by Maxwell et al. \cite{Maxwell77a}.

In recent years, some authors made a strong case for the condensation
of kaons in the cores of neutron stars (see, for instance,
\cite{Kaplan86a,Brown87a,Lee95a}). The critical density beyond which 
this process may
take place is uncertain.  Estimates lie in the range of $3-5\,n_0$
\cite{Thorsson94a,Brown88a,Muto92a}.  The emissivities of the
quasi-particle direct Urca processes,
\begin{equation}
  \mathrm{n(K)} \rightarrow \mathrm{p(K)} + \mathrm{e}^-
  +\bar{\nu}_\mathrm{e} \qquad \mbox{and} \qquad \mathrm{p(K)} +
  \mathrm{e}^- \rightarrow \mathrm{n(K)} + \nu_\mathrm{e} \; ,
\label{eq:kmes1}
\end{equation} are reduced, relative to the uncondensed case, by a
factor of $\cos^2(\theta/2)$ \cite{Thorsson95a}. Additionally to
(\ref{eq:kmes1}) there are two more processes,
\begin{eqnarray}
  \mathrm{n(K)} \rightarrow \mathrm{n(K)} + \mathrm{e}^-
  +\bar{\nu}_\mathrm{e} \qquad &\mbox{and} &\qquad \mathrm{n(K)} +
  \mathrm{e}^- \rightarrow \mathrm{n(K)} + \nu_\mathrm{e}\; , \\ 
  \mathrm{p(K)} \rightarrow \mathrm{p(K)} + \mathrm{e}^-
  +\bar{\nu}_\mathrm{e} \qquad &\mbox{and} &\qquad \mathrm{p(K)} +
  \mathrm{e}^- \rightarrow \mathrm{p(K)} + \nu_\mathrm{e} \; ,
\end{eqnarray} whose emissivities were calculated by Thorsson et al.
\cite{Thorsson95a}. The density dependence of the mixing angle
$\theta$ and the value of the critical density, $n_\mathrm{c}=4.18
\,n_0$ are taken from \cite{Thorsson94a}.

%
\subsection{Quark processes} \label{subsec:emisquark}

The possible existence of quark matter in the cores of neutron stars
has already been outlined in sect. \ref{ssec:eos:comparison}.  The
latest calculations on this topic predict an onset of the transition
of confined hadronic matter into quark matter at $n\approx 1.6\, n_0$.
(This is not to be confused with the possibility of absolutely stable
strange matter, to which we will return in sect.
\ref{sec:strangestars}).

The following weak processes are important in quark matter consisting
of up, down and strange quarks,
\begin{eqnarray}
  \mathrm{d} \rightarrow \mathrm{u} + \mathrm{e}^- +
  \bar{\nu}_\mathrm{e} \qquad &\mbox{and} &\qquad \mathrm{u} +
  \mathrm{e}^- \rightarrow \mathrm{d} + \nu_\mathrm{e}\; , 
\label{eq:quark1} \\
  \mathrm{s} \rightarrow \mathrm{u} + \mathrm{e}^- +
  \bar{\nu}_\mathrm{e} \qquad &\mbox{and} &\qquad \mathrm{u} +
  \mathrm{e}^- \rightarrow \mathrm{s} + \nu_\mathrm{e}\; .
\label{eq:quark2}
\end{eqnarray} We shall use the neutrino emission rates given by Duncan,
Shapiro, and Wasserman \cite{Duncan83} to simulate the cooling history
of neutron stars containing a mixed phase of baryons and quarks in
their interior (hybrid stars) and strange stars.

\section{Results and comparison with observed data}\label{sec:res}

\subsection{Test of the numerical code}

\begin{figure}[tbp] \centering
  \psfig{figure=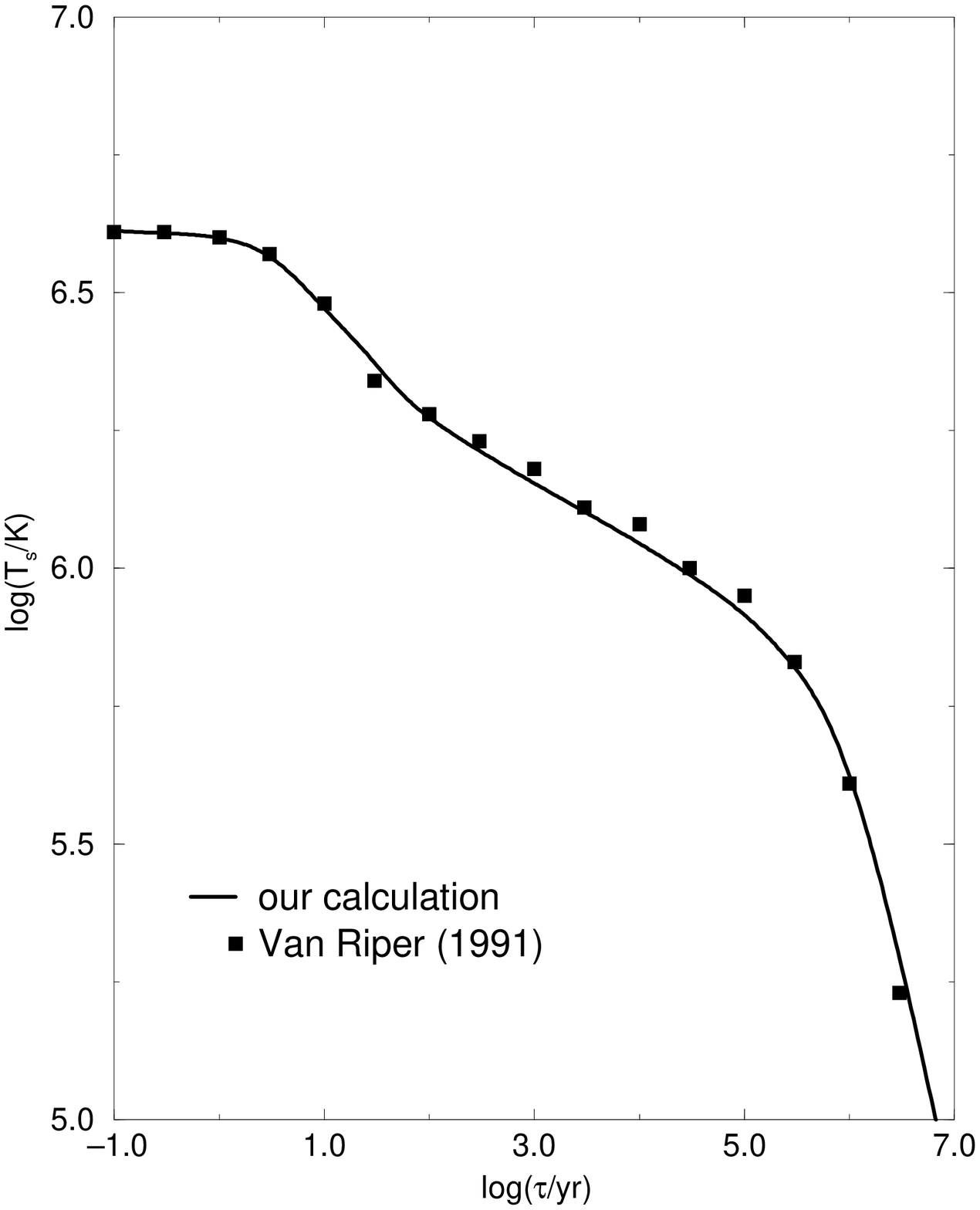,width=7cm}
  \capt{Comparison of the computed surface temperature versus age with
  the results of Van Riper \cite{VanRiper91}. The parameters are:
  $M=1.4M_{\odot}$, BPS EOS \cite{Baym71}, no superfluidity and no
  enhanced cooling processes are included.}
 \label{fig:vanrip}
\end{figure}

We tested the correctness of our numerical code by means of comparing
the outcome of several cooling calculations with those obtained by
other authors. As an example, in fig. \ref{fig:vanrip} we compare
our surface temperatures, obtained for a particular model for the
equation of state, with the corresponding results obtained by Van
Riper \cite{VanRiper91}. The agreement is extremely good for all star ages.
To achieve this agreement, however, we had to reduce the core's
thermal conductivity, which was not specified by Van Riper, from the
value given in \cite{Gnedin95a} by a factor of 40. Otherwise deviations
of about $\sim 20\%$ were obtained at star ages between $\tau \sim 1$
to 10~years. The perfect agreement at later stages however is not
influenced by this.

%
\subsection{Body of observed cooling data}\label{sec:observations}

{}From the X-ray observatories Einstein, EXOSAT, and ROSAT the X-ray
spectra of at least 14 pulsars have been observed to date. Unfortunately, the
data are not always sufficient to extract the effective surface
temperature of the corresponding pulsar. One can classify the spectra
into three categories \cite{Oegelman95a}  (compare with table
\ref{tab:observations}):
\begin{table}
  \centering \capt{Survey of observed pulsar ages and luminosities, denoted
    $\tau$ and $L$, respectively\label{tab:observations}}
  \smallskip
\begin{tabular}{ccccc}
  Pulsar & Name & $\log\tau$ [yr] & $\log L$ [erg/s] & Reference \\
  \hline \hline
  \multicolumn{5}{c}{\small not enough data available for spectral analysis} \\
  1706-44 & & 4.25 & $32.8\pm 0.7$ & \cite{Becker92a} \\
  1823-13 & & 4.50 & $33.2\pm 0.6$ & \cite{Finley93b} \\
  2334+61 & & 4.61 & $33.1\pm 0.4$ & \cite{Becker93b} \\
  \hline
  \multicolumn{5}{c}{\small power-law-type spectra or spectra with only a 
	high temperature component} \\
  0531+21 & Crab & 3.09 & $35.5\pm 0.3$ & \cite{Becker95a}\\
  1509-58 & SNR MSH 15-52 & 3.19 & $33.6\pm 0.4$ 
        & \cite{Seward83a,Trussoni90a} \\
  0540-69 & & 3.22 & $36.2\pm 0.2$ & \cite{Finley93a} \\
  1951+32 & SNR CTB 80 & 5.02 & $33.8\pm 0.5$ & \cite{SafiHarb95a} \\
  1929+10 & & 6.49 & $28.9\pm 0.5$ & \cite{Oegelman95a,Yancopoulos93} \\
  0950+08 & & 7.24 & $29.6\pm 1.0$ & \cite{Seward88a} \\
  J0437-47 & & 8.88 & $30.6\pm 0.4$ & \cite{Becker93c} \\
  \hline
  \multicolumn{5}{c}{\small spectrum dominated by a soft component} \\
  0833-45 & Vela & 4.05 & $32.9\pm 0.2$ & \cite{Oegelman93a} \\
  0656+14 & & 5.04 & $32.6\pm 0.3$ & \cite{Finley92a} \\
  0630+18 & Geminga & 5.51 & $31.8\pm 0.4$ & \cite{Halpern93a} \\
  1055-52 & & 5.73 & $33.0\pm 0.6$ & \cite{Oegelman93b} \\
  \hline
\end{tabular} \comments{The age is determined by the spin down of the
  pulsar, the luminosities by spectral fits or the totally detected
  photon flux, depending on the categories (i) to (iii) listed in the
  text.}
\end{table}

\begin{enumerate}
\item The photon yields of three pulsars (PSR's 1706-44, 1823-13, and
  2334+61) contain too few photons in order to perform spectral fits.
  The luminosities are calculated by using the totally detected photon
  flux.  These pulsars are marked with triangles in the figures.
\item The spectra of seven pulsars, including the crab pulsar (PSR
  0531+21), can only be fitted by a power-law type spectrum or a
  blackbody spectrum with a very high effective temperature, and an
  effective area much smaller than the neutron star surface. Their
  X-ray emission is predominated by magnetospheric emission.
  Therefore, their temperatures, determined from the spectral fits, are
  probably too high. Theses pulsars are marked with solid dots.
\item Finally, there are four pulsars (i.e., 0833-45 (Vela), 0656+14,
  0630+18 (Ge-minga), and 1055-52) that allow two-component spectral
  fits, where the softer blackbody components corresponds to the
  actual surface emission of the neutron star and the harder to some
  magnetospheric emission.  These pulsars are marked with squares.
\end{enumerate}

When discussing spectra of neutron stars one should keep in mind that
neutron stars are not black bodies, because hydrogen and helium in
their atmospheres modify the blackbody spectrum. Besides, strong
magnetic fields can also affect the surface emission.  With respect to
the luminosity determinations we note that the luminosities determined
via hydrogen- or helium-atmospheric models are generally lower by a
factor of about three than those determined via blackbody spectral
fits. The inverse is true for the temperature which is smaller by a
factor of about two. An uncertainty in temperature by a factor of two
varies the luminosity, according to the Stefan-Boltzmann law
($L\propto T^4$), by a factor of 16. It is thus not appropriate to
perform a comparison with surface temperatures of stars
\cite{Oegelman95a}.  Instead one should be using the luminosities, as
is the case here.

\subsection{Standard cooling calculations} \label{sec:standard}

\begin{figure}[tbp] \centering 
  \psfig{figure=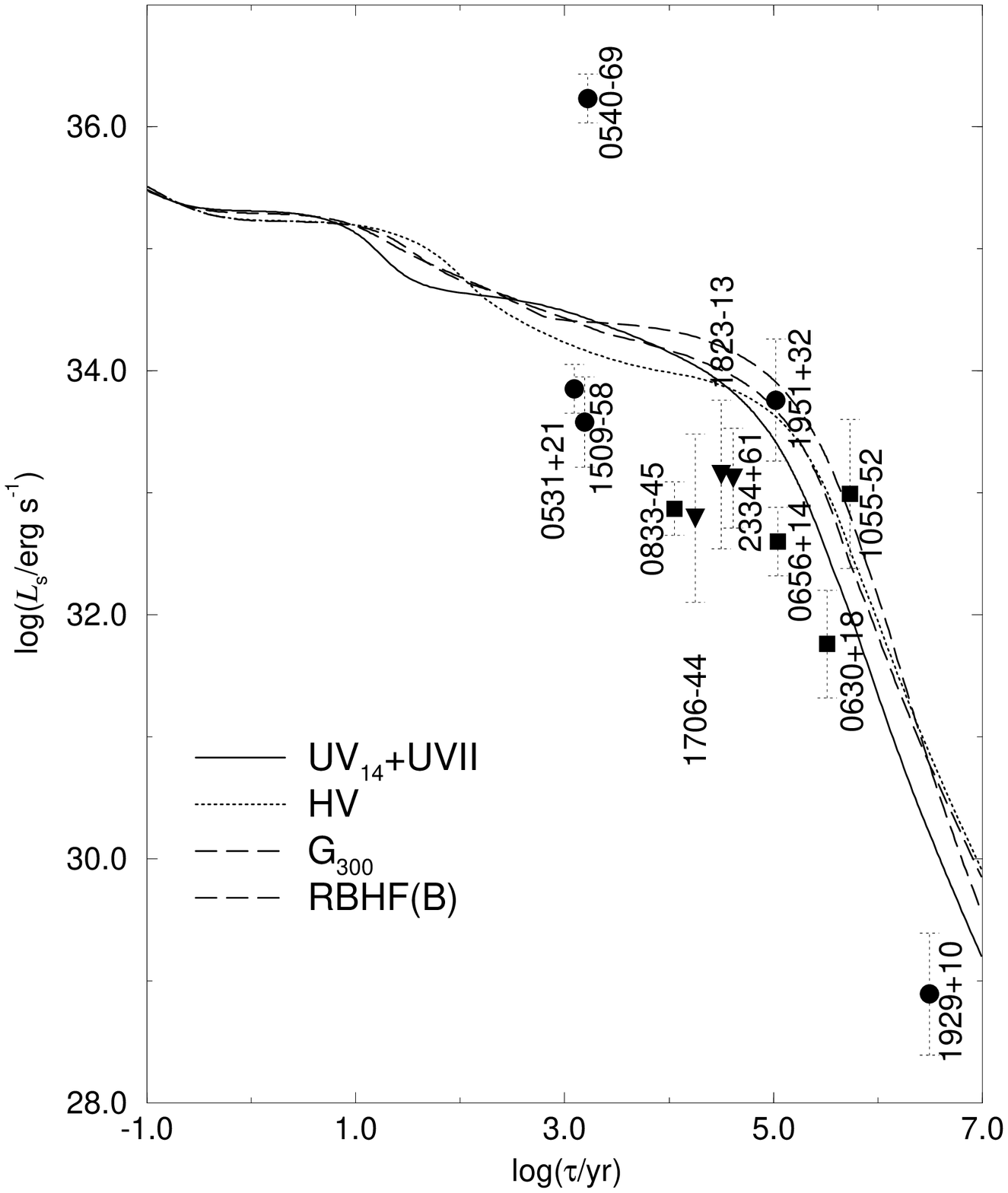,width=7cm} 
\capt{Standard cooling behavior of a neutron star with mass $M=1.4M_\odot$
    for several different models for the EOS.}
  \label{fig:standard}
\end{figure}

Figure \ref{fig:standard} shows the cooling history of a neutron star
with $M=1.4M_\odot$ that cools down via the standard scenario only,
i.e., fast cooling processes are not allowed. The different curves
correspond to different models for the equation of state.  One sees
that the standard cooling behavior depends only weakly on the equation
of state. The \UVU~ model cools somewhat faster at later times than
the other two models. This is due to the rather soft behavior of the
\UVU~ equation of state at intermediate nuclear densities
(fig. \ref{fig:eos}), which leads to neutron stars with small radii
(figs. \ref{fig:rho:r:0.5}--\ref{fig:rho:r:1.8}).  Obviously, most of
the observed data are not in agreement with the assumption of standard
cooling.  Specifically Vela's (0833-45) luminosity lies considerably
below the computed curves.  The data of PSR 0630+18 (Geminga) and a
few others are in relatively good agreement, however.\footnote{Note
that there is a considerable uncertainty in the ages of pulsars that
are derived from the braking index, $n$, which, due to lack of
information, usually is taken to be constant during a pulsar's
spin-down. Moreover a canonical value of $n=3$
\cite{Shapiro83}, which corresponds to a slow-down by emission of pure
magnetic dipole radiation, is assumed.  This too constitutes a rather
crude approximation.}

\subsection{Enhanced cooling via direct Urca process}\label{ssec:edurca}

\begin{figure}[tbp] \centering 
  \psfig{figure=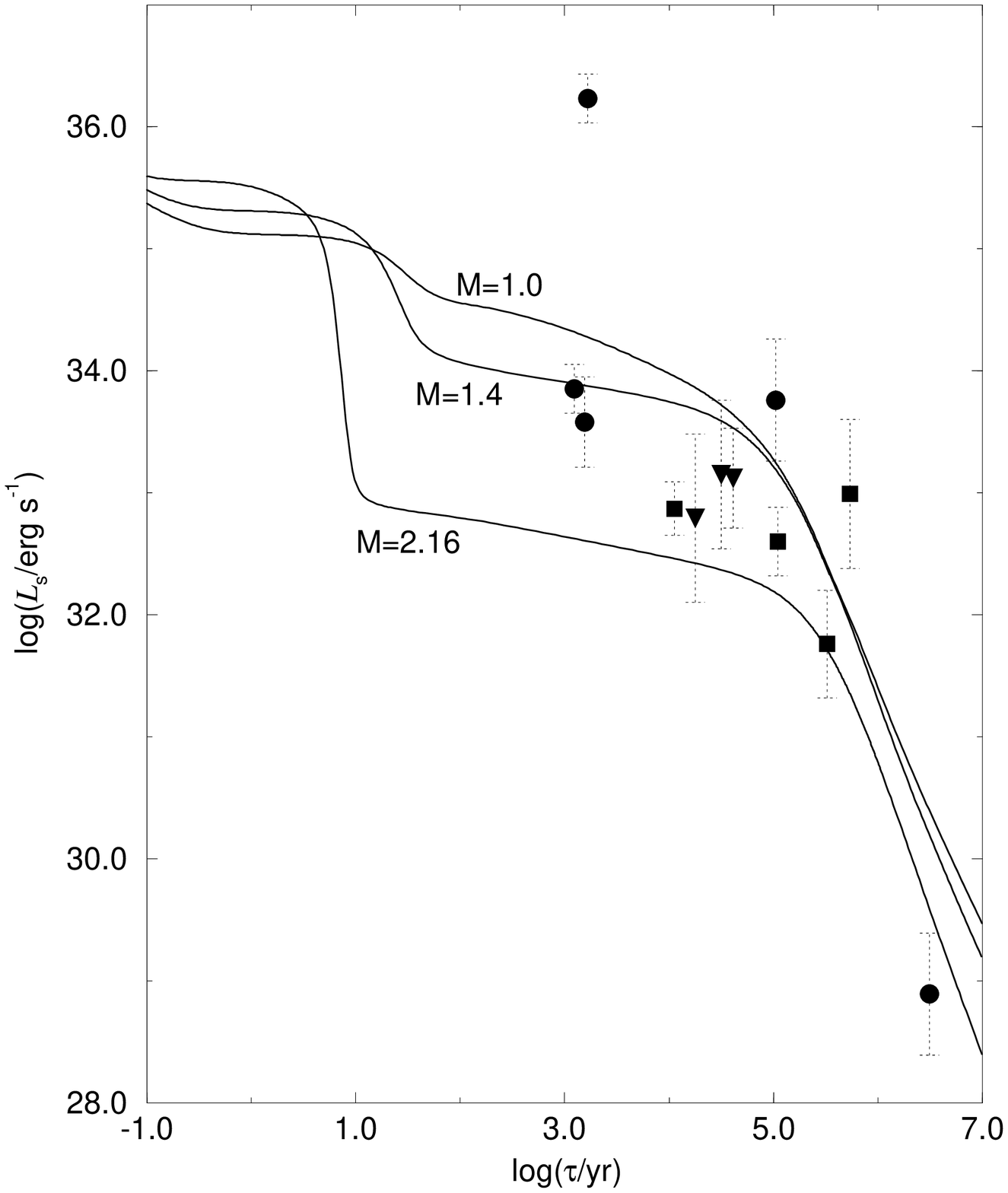,width=7cm}
  \capt{Enhanced cooling of neutron stars via direct Urca process. The
  underlying EOS is \UVU. The observed data are labeled in fig.
  \protect{\ref{fig:standard}}.}
  \label{fig:dr.uvu}
\end{figure}

\begin{figure}[tbp] \centering 
  \psfig{figure=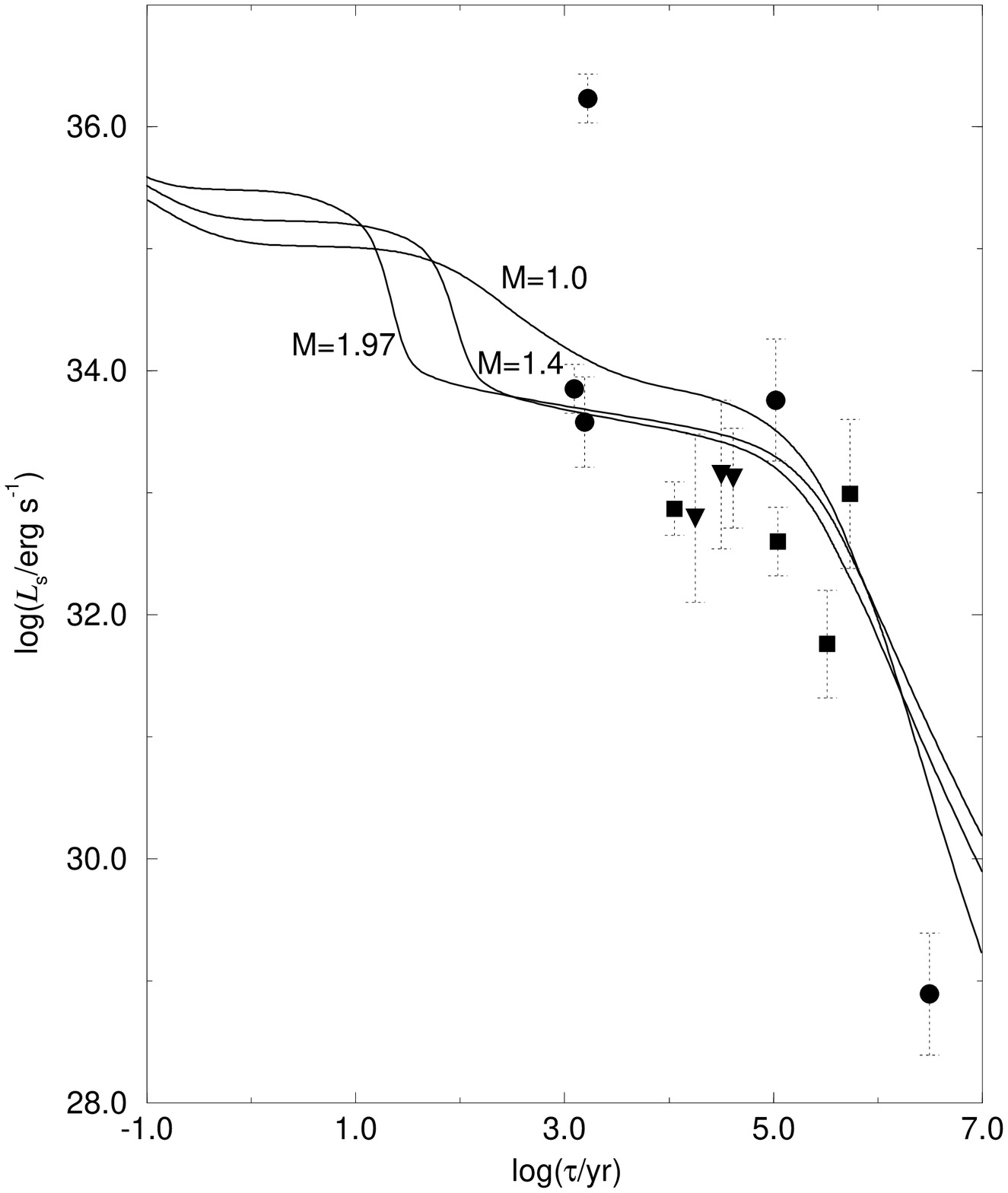,width=7cm} 
\capt{Same as figure \protect{\ref{fig:dr.uvu}} but for HV. The observed
  data are labeled in fig. \protect{\ref{fig:standard}}.}
  \label{fig:dr.hv}
\end{figure}

\begin{figure}[tbp] \centering 
  \psfig{figure=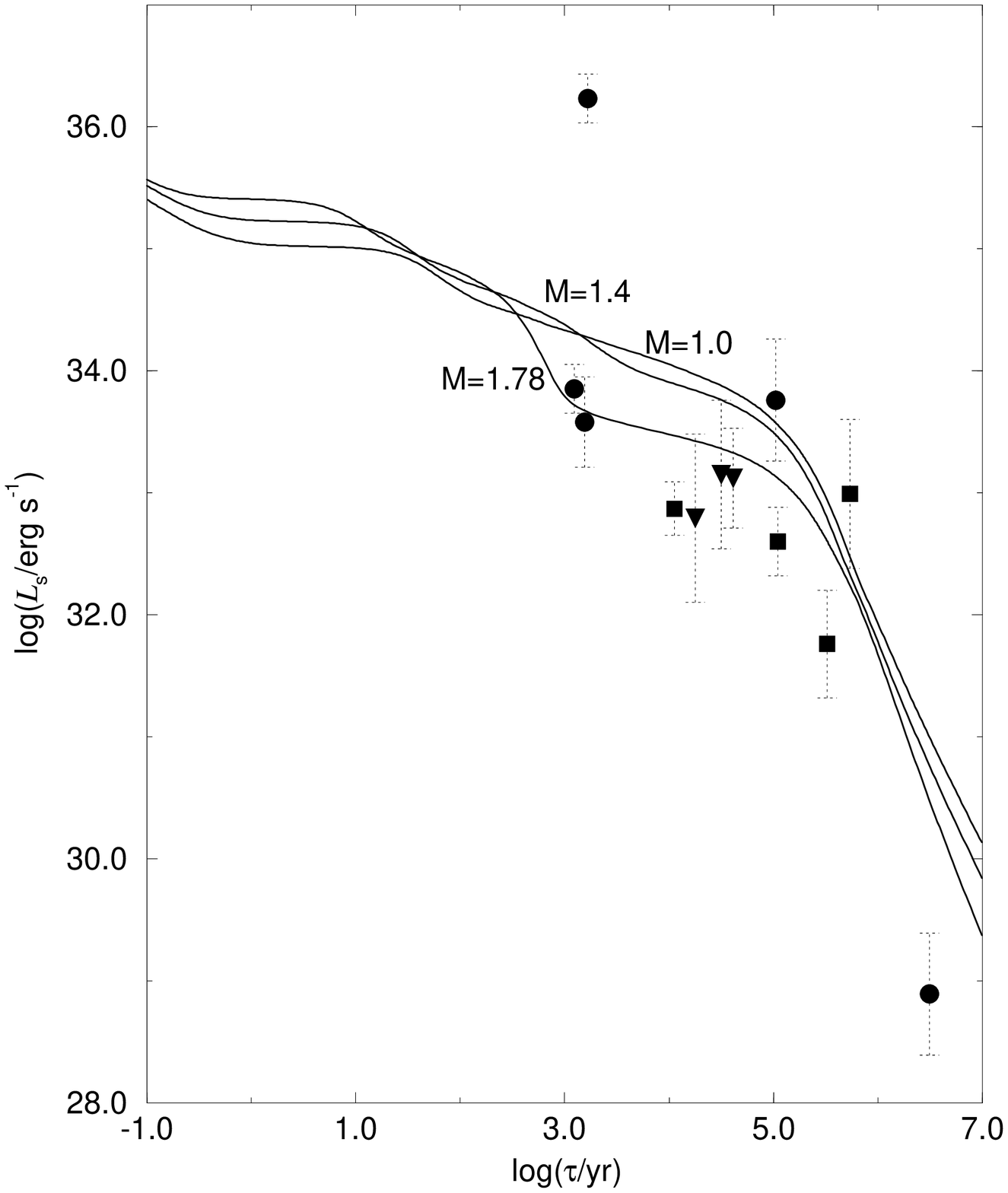,width=7cm} 
\capt{Same as fig. \protect{\ref{fig:dr.uvu}} but for \glen. The observed
  data are labeled in figure \protect{\ref{fig:standard}}.}
  \label{fig:dr.g300}
\end{figure}

As mentioned above, one way by means of which neutron stars can cool
very efficiently is the direct Urca process 
\cite{Page92:b,Gnedin93:a,Umeda94:ab}. This is demonstrated
graphically in figs. \ref{fig:dr.uvu}--\ref{fig:dr.g300} for the
equations of state \UVU, HV, and \glen. The proton fraction in the
stars with $M=1.0M_\odot$ is for all three cases below the threshold
density for the direct Urca process. Hence these stars cool via the
standard mechanism, the modified Urca process.  The cooling behavior
of all other stars at early stages is dominated by the direct Urca
process.  The sudden drop in luminosity at ages of several hundred
years has its origin in the fact that the core cools down much faster
through the enhanced cooling process than the crust. This causes a
temperature inversion in young stars. After several hundred years,
depending on the thickness of the crust, the ``\emph{cooling wave}'',
built up by the temperature gradient, reaches the star's surface which
leads to the sudden luminosity drops. (An estimate of the time needed
by the cooling wave to reach the surface can be found in
\cite{Lattimer94a}.)  The drop of luminosity increases with massive,
since the region inside the core of a star where the direct Urca
process is possible grows with density.  The drop is most strongly
pronounced for \UVU. This has its origin in the \sfp~
neutron-superfluid which does not extent into the central core.

\begin{figure}[tbp] \centering 
  \psfig{figure=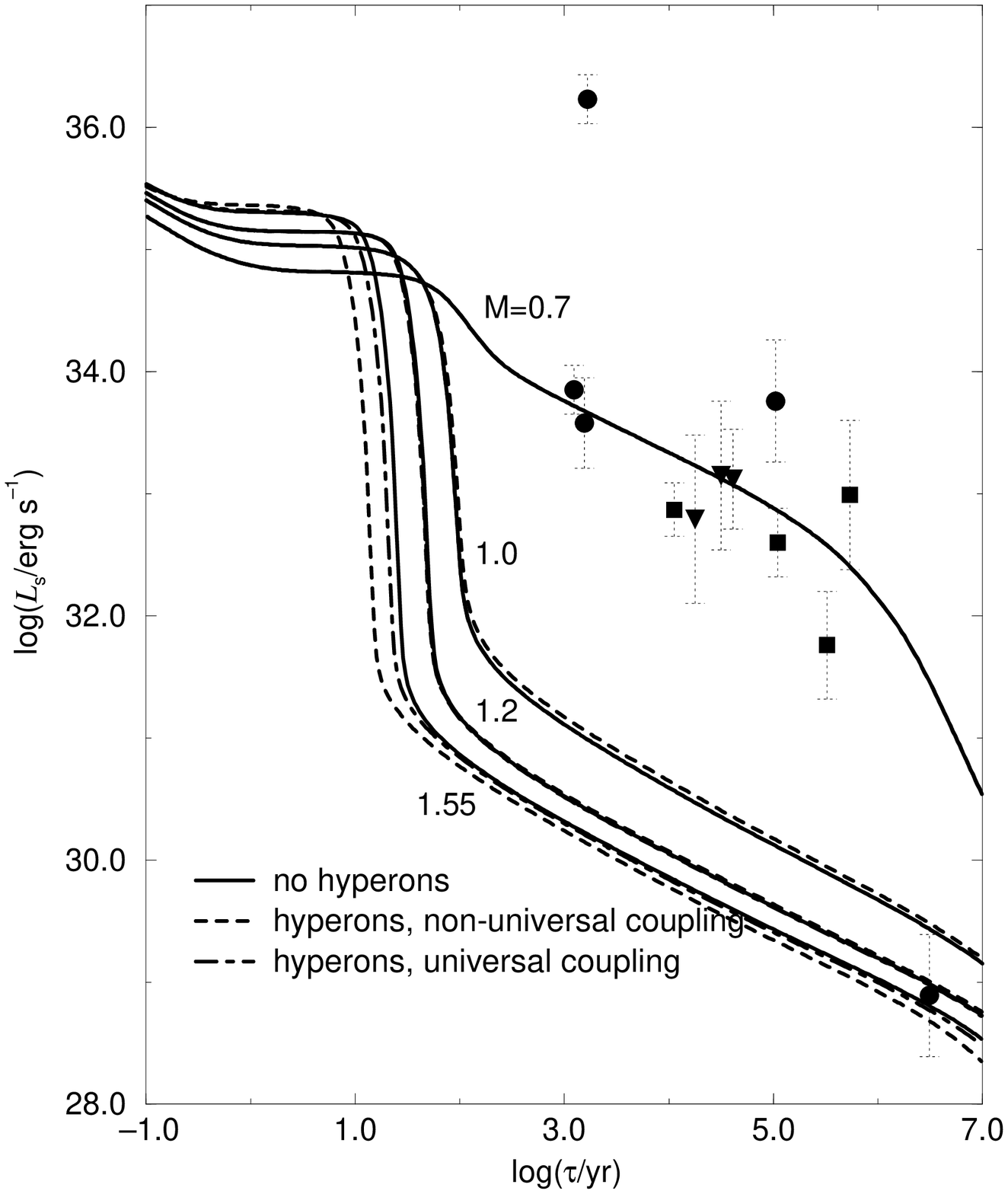,width=7cm}
  \capt{Enhanced cooling of neutron stars due to direct Urca among
nucleons, but for varying hyperon coupling strengths (see text).  The
underlying EOS is \KM. The observed data are labeled in
fig. \protect{\ref{fig:standard}}.}  \label{fig:universal}
\end{figure}

Changes in the hyperon couplings strengths in matter influence the
cooling curves only very little, as can be seen in
fig. \ref{fig:universal} for neutron stars of several different
masses. (Note that this may be drastically different if the
direct Urca process among hyperons occurs is dense stellar matter
\cite{Prakash92,Haensel94:a}.) The star with $M=0.7\,M_\odot$ does 
not contain hyperons, since its core is not dense enough to reach the
threshold density for hyperon population. Hence there is only one
cooling curve in this case. The situation is different for the heavier
stars shown in this figure, for which the hyperon concentration
increases with star mass
\cite{Glendenning85a}. The hyperons are coupled to the meson
fields either with the same strengths as the nucleons, i.e.,
$x_\sigma=x_\omega=x_\rho=1$ (cf. Eq. (\ref{eq:ghyp})), or are chosen
such that mutual compatibility between the $\Lambda$ binding energy in
saturated nuclear matter, neutron-star masses, and experimental data
on hypernuclear levels is achieved
(cf. sect. \ref{ssec:eos:comparison}). As for the other figures of
this section, the direct Urca process among the nucleons is
included, and no superfluidity is taken into account.

\subsection{Enhanced cooling via pion condensation}

\begin{figure}[tbp] \centering 
  \psfig{figure=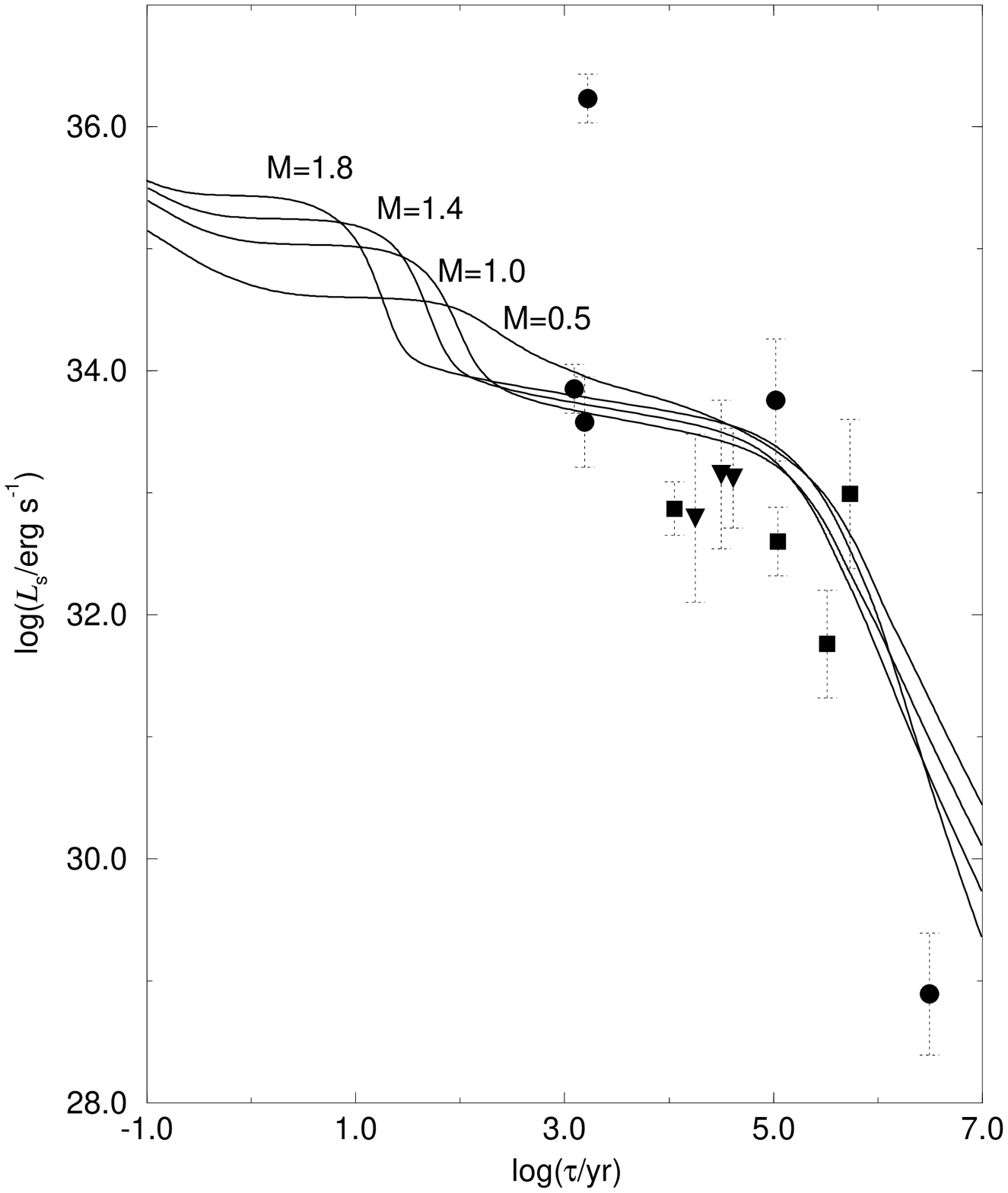,width=7cm}
  \capt{Enhanced cooling of neutron stars due to pion condensation.
  The underlying EOS is \glenpi. The observed data are labeled in
  fig. \protect{\ref{fig:standard}}.}
  \label{fig:pion.g300}
\end{figure}

The impact of pion condensation on the cooling of neutron stars 
\cite{Umeda94:ab} is demonstrated in fig. \ref{fig:pion.g300}. 
comparison with fig. \ref{fig:dr.g300} shows that the cooling behavior
of such stars is qualitatively similar to the one of stars which cool
via the direct Urca process. The threshold density of the pion
condensate lies at $3.6 \times 10^{14}~\gccm$, just slightly above
normal nuclear matter saturation density.  Hence all stars displayed
in fig.
\ref{fig:pion.g300}, with the exception of $M=0.5M_\odot$, possess a
pion condensate and therefore show enhanced cooling triggered by the
condensate.

\subsection{Enhanced cooling via kaon condensation}\label{ssec:kaon}

\begin{figure}[tbp] \centering 
  \psfig{figure=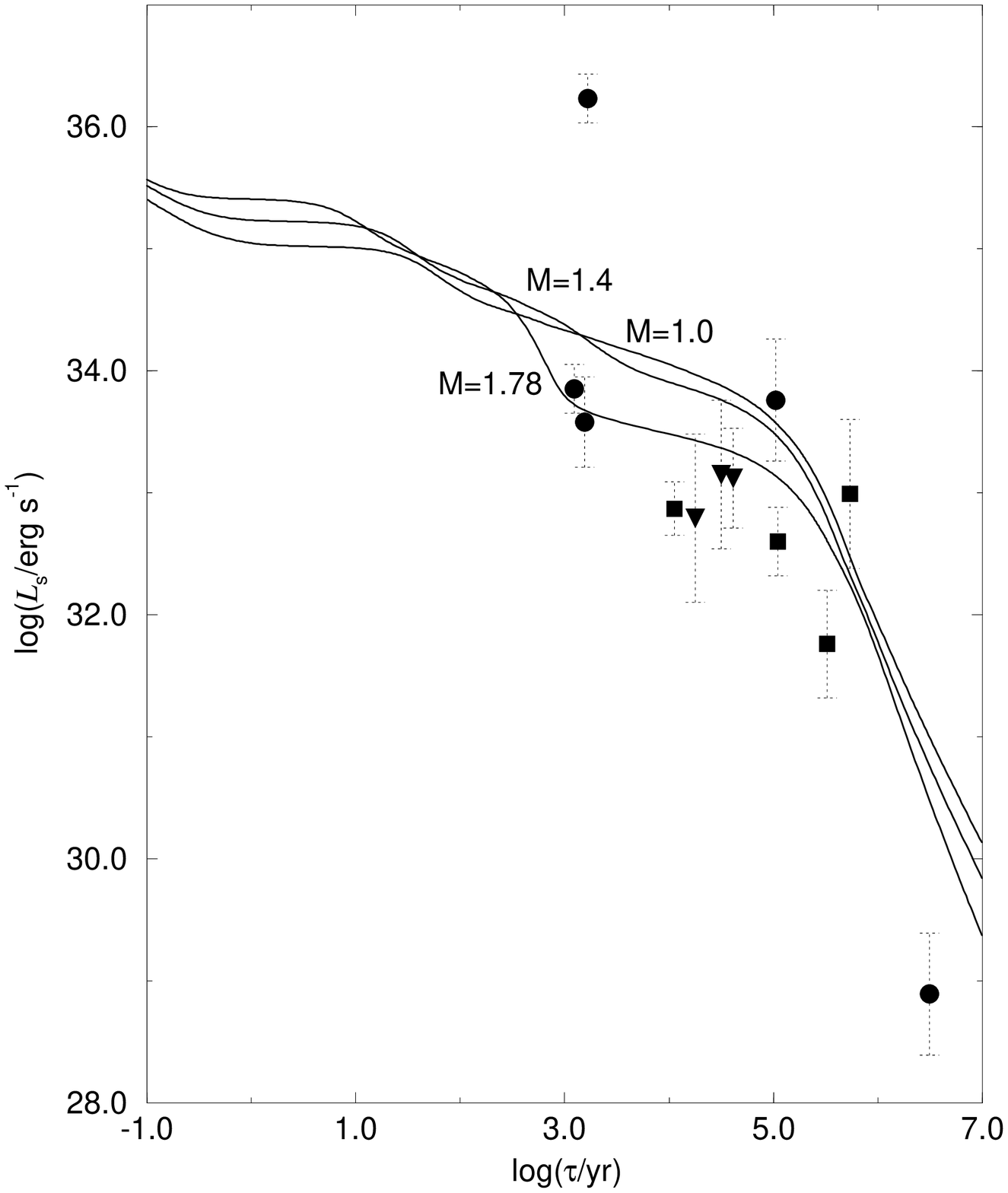,width=7cm}
  \capt{Enhanced cooling of neutron stars due to kaon condensation.
  The underlying EOS is \glen. The observed data are
  labeled in fig. \protect{\ref{fig:standard}}.}
  \label{fig:kaon.g300}
\end{figure}

The cooling behavior of kaon-condensed stars (see also 
\cite{Page95,Umeda94:ab,Page90:a,Page93:a,Page94:a})
is qualitatively similar to those of pion-condensed stars. This is
graphically illustrated in fig. \ref{fig:kaon.g300} for the
\glen~equation of state. There is however a significant quantitative
difference between both types of condensation which concerns the
threshold densities. It amounts for kaon condensation about $4.8\,n_0$
\cite{Thorsson94a}, which is by a factor of $\sim 2.5$ larger than the
threshold density established for pion condensation, $\sim 1.3\,n_0$
\cite{Glendenning85a}. A density of $4.8\,n_0$ is reached only in the
most massive star model, $M=1.78\,M_\odot$, constructed for
\glen. Hence the other two models cool via the modified Urca process
only.  For the purpose of comparison, the central density of the
$M=1.4\,M_\odot$ neutron star amounts $2.4\,n_0$ which lies almost by
a factor of two below the kaon threshold. The situation would be
different for softer models for the equation of state, which lead to
higher central neutron star densities. For the broad collection of
modern equations of state of ref.  \cite{Weber91} we find central
densities in $M\sim 1.4\,M_\odot$ stars that range from about
$2.3\,n_0$ to $4.2\,n_0$. The latter value comes much closer to the
critical value for kaon condensation derived in \cite{Thorsson94a},
which, of course, is model dependent too and thus not accurately
known. Clearly, the smaller the threshold density the faster the cooling
of compact stars via kaon condensation will be.

\subsection{Enhanced cooling due to the presence of quark matter in 
  the cores of neutron stars}\label{ssec:ecquark}

\begin{figure}[tbp] \centering 
  \psfig{figure=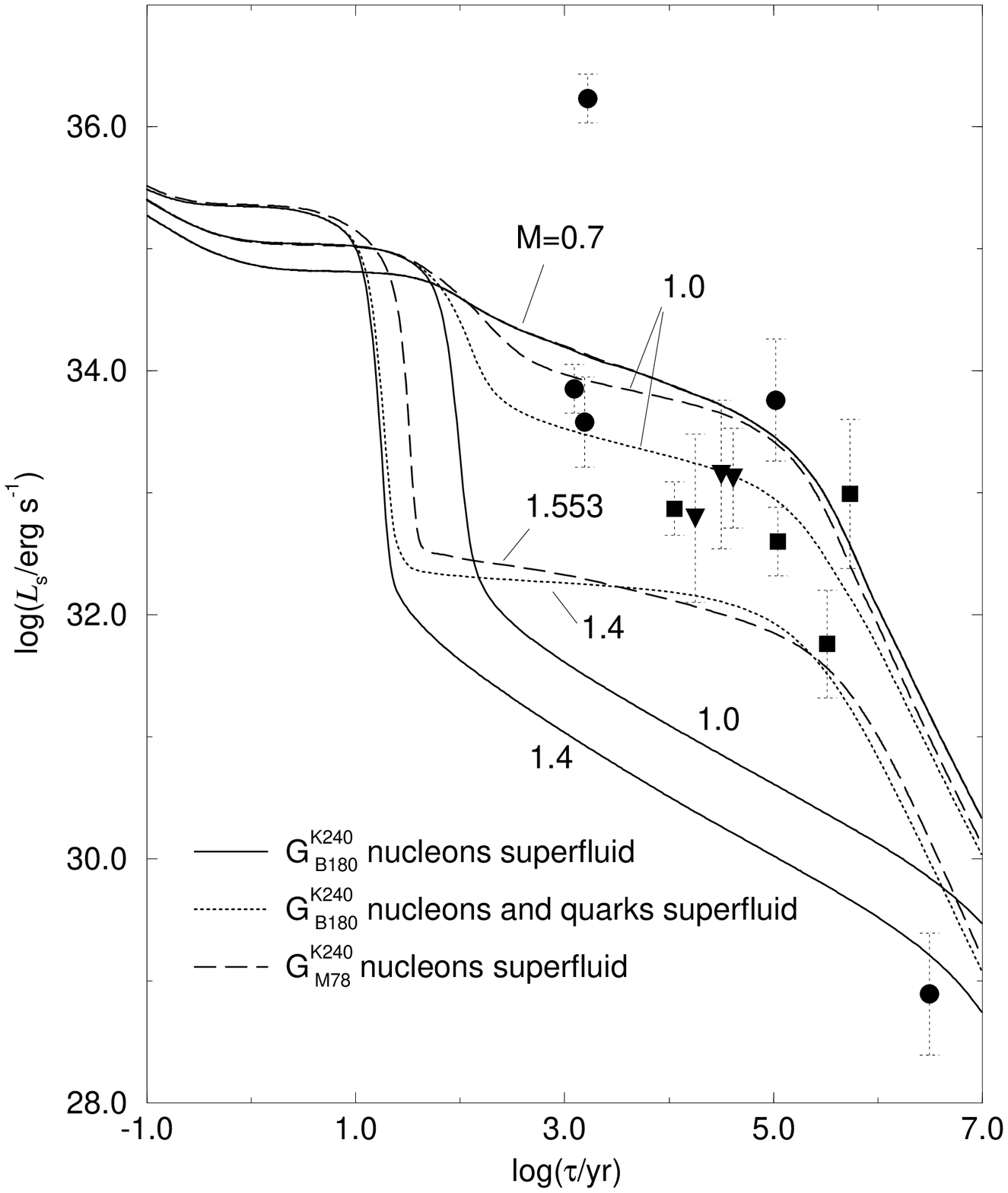,width=7cm}
  \capt{Enhanced cooling of neutron stars due to an admixture of quark
  matter in their cores (\KB~EOS). Curves corresponding to the
  \KM~EOS are without quarks. Enhanced cooling via direct Urca is
  taken into account too. The observed data are labeled in fig.
  \protect{\ref{fig:standard}}.}
  \label{fig:k240b180}
\end{figure}

As discussed before, modern investigations of the possible transition
of baryon matter into quark matter predict transition densities less
than twice normal nuclear matter density. Such densities are easily
reached in neutron stars with masses $M\approx 1.4 M_\odot$
\cite{Weber91} (see also discussion at the end of section
\ref{ssec:kaon}).  To simulate the cooling history of such (hybrid)
stars we apply Glendenning's \KB~equation of state, whose specific
properties have been discussed in connection with table
\ref{tab:eos2}. The weak reaction processes of the quarks are those
given in Equations (\ref{eq:quark1}) and (\ref{eq:quark2}).  They
speed up the energy loss of a star considerably, as is illustrated in
fig.  \ref{fig:k240b180}, where the solid curves correspond to
enhanced cooling caused by the quarks (see also \cite{VanRiper91}).
The dashed curves refer to stars with no quarks in their cores,
which therefore cool more slowly.  Note that the density in the light
stars with $M<1\,M_\odot$ is not high enough to reach the transition
into quark matter. Enhanced quark cooling is therefore not possible
for these stars. Finally, if the quarks should form a superfluid, the
emission of neutrinos is suppressed and cooling proceeds at a much
slower pace (dotted curves).  Bailin and Love have stated that quarks
might become superfluid \cite{Bailin84}.  Unfortunately, there exists
-- as far as we know -- no estimation of the gap energy of superfluid
quark matter.  Therefore, in order to simulated this effect, we simply
assumed a plausible value of $\Delta_\mathrm{sf}=0.1$~MeV for the
\sfs~ pairing state of quarks, which was taken to be constant over the
whole density range.

\subsection{Strange quark matter stars}

\begin{figure}[tbp] \centering 
  \psfig{figure=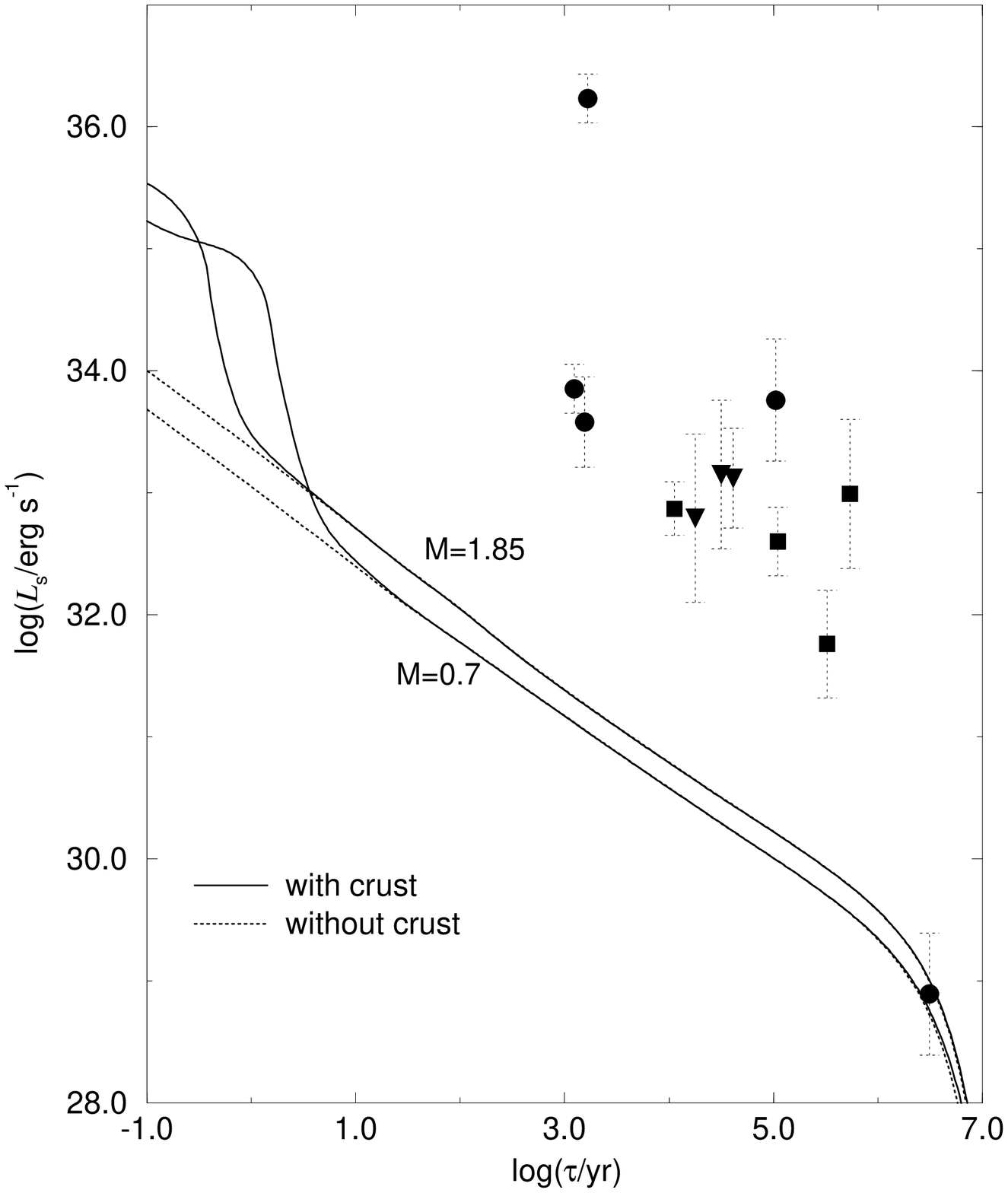,width=7cm} 
\capt{Cooling behavior of strange quark matter stars of masses
  $M=0.7\,M_\odot$ and $M=1.85\,M_\odot$ with and without crust. The
  quarks are treated as non-superfluid. The observed data are
  labeled in fig.  \protect{\ref{fig:standard}}.}
  \label{fig:sm.nosf}
\end{figure}

\begin{figure}[tbp] \centering 
  \psfig{figure=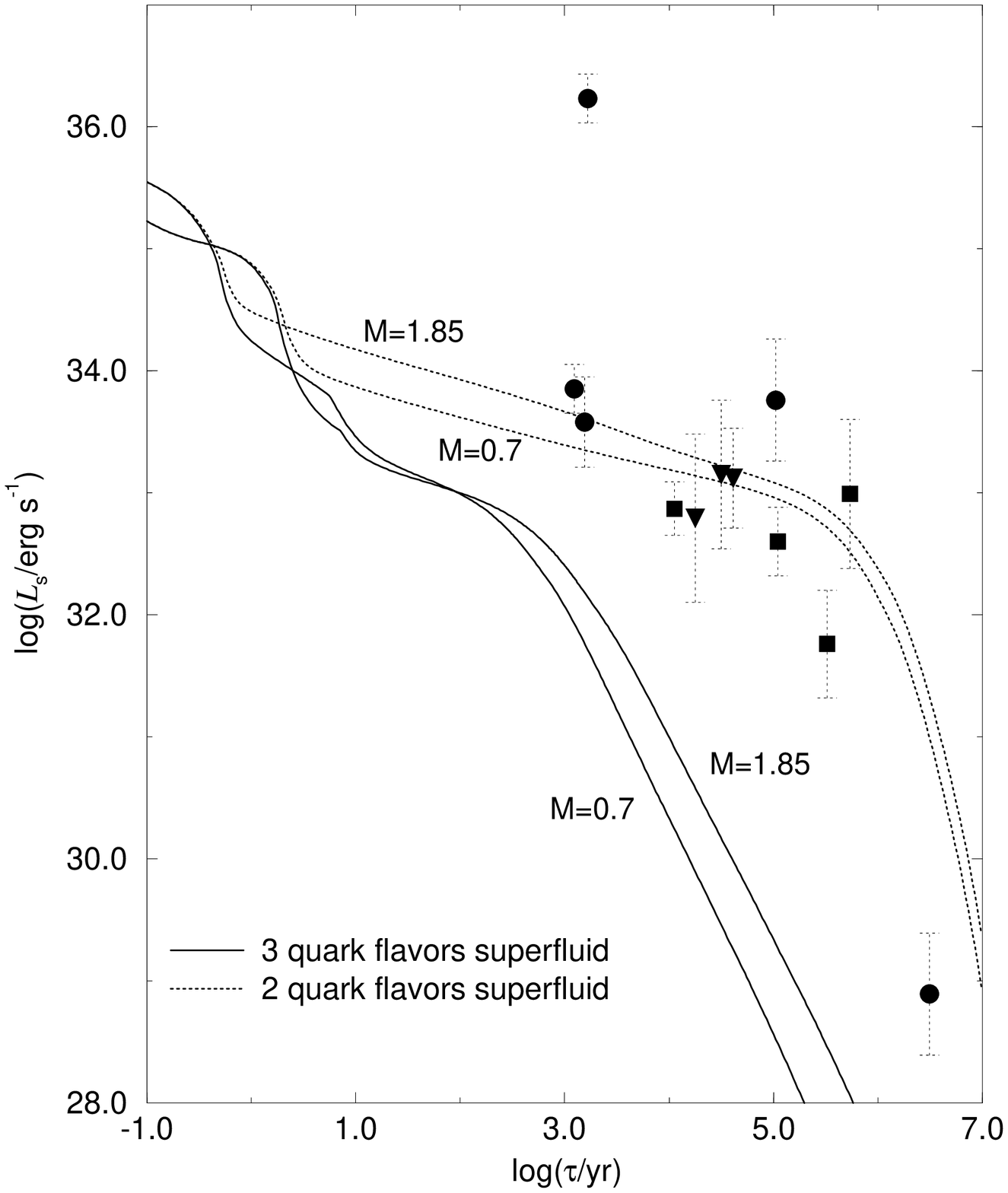,width=7cm} 
\capt{Cooling behavior of strange quark matter stars with crust. The
  masses are $M=0.7\,M_\odot$ and $M=1.85\,M_\odot$. The quarks are
  treated as superfluid particles.  The heat capacity is reduced by
  an exponential factor in one model, while in the other model it is
  kept unchanged. The observed data are labeled in fig.
  \protect{\ref{fig:standard}}.}
  \label{fig:sm.sf}
\end{figure}

We restrict ourself to determine the cooling behavior of strange stars
which are either bare or possess the densest possible hadronic crust
that can be carried by such objects (inner crust density equal to
neutron drip density). Figure \ref{fig:sm.nosf} reveals that, no
matter how thick the crust, such stars cool very rapidly
\cite{Page92:a}. Among the body of observed luminosities, there is
presently one data point that would be consistent with the assumption
that the underlying object is a strange star. This interpretation is
not stringent of course, because some of the enhanced cooling
mechanisms studied above might do as well.  If one treats the quarks
as superfluid, neutrino production ceases and therefore cooling is
considerably slowed down.  To simulate this effect qualitatively we
assumed a density-independent energy gap of
$\Delta_\mathrm{sf}=0.1$~MeV. The numerical outcome of such a
simulation is exhibited in fig. \ref{fig:sm.sf}.  One sees that in
this case the cooling behavior of strange stars is similar to the
enhanced cooling behavior of neutron stars (compare figs.
\ref{fig:dr.uvu}--\ref{fig:k240b180}). Hence, with the exception of
PSR 0540-69 practically all observed pulsar luminosities can now be
reproduced too. A quantitative clarification of the question as to
whether or not quark matter indeed becomes superfluid, and if so, at
which densities, attains particular interest from astrophysics too.
There is also the possibility that the quark flavors
become superfluid at different densities. Consequently the heat
capacity would be somewhat smaller. To simulate this possibility we
reduce the heat capacity by an (arbitrary) factor of three. The
consequences for the cooling behavior are shown by the dotted curves
in \ref{fig:sm.sf}.

In closing this section, we mention that one of the main differences
between the cooling behavior of neutron stars and strange quark matter
stars concerns the time needed for the cooling wave to reach the
surface.  This time is much shorter for strange quark matter stars
because their crusts have a thickness of only $\sim 0.2$~km as opposed
to $\sim 2$~km for stars with $M\sim 1.4M_\odot$.

\section{Conclusions}\label{sec:con}

We have already pointed out in section \ref{sec:observations} that the
observed cooling data of pulsars marked with squares in our figures
appear to be the most significant ones. Primarily these data should be
reproduced by theoretical cooling calculations. As we have seen above
(section \ref{sec:standard}) the luminosities of a few pulsars, like
1951+32 and 1055-52, appear to be compatible with the standard cooling
scenarios.  Depending on the model employed for the equation of state,
this may also be the case for Geminga and pulsars 0656+14, 2334+61
(cf. fig. \ref{fig:standard}). Others, however, like the famous Vela
pulsar (PSR 0833-45) and PSR's 1706-44, 1929+10, which have rather low
observed luminosities, may require the introduction of so-called fast
(or enhanced) cooling processes, which are characterized by higher
neutrino emission rates by means of which stars cool more rapidly. A
final decision on this issue -- standard versus enhanced cooling
\cite{Oegelman95a} -- is presently hampered by the uncertainties in a
number of quantities, including the modified Urca neutrino-emission
rate, superfluid gaps, equation of state, and mass and age of the
star. Nevertheless, subject to the enormous progress that is being
made presently in exploring the behavior of superdense matter
\cite{Greiner94a}, one may feel confident that this will change in the
foreseeable future.
 
\begin{figure}[tbp] \centering 
  \psfig{figure=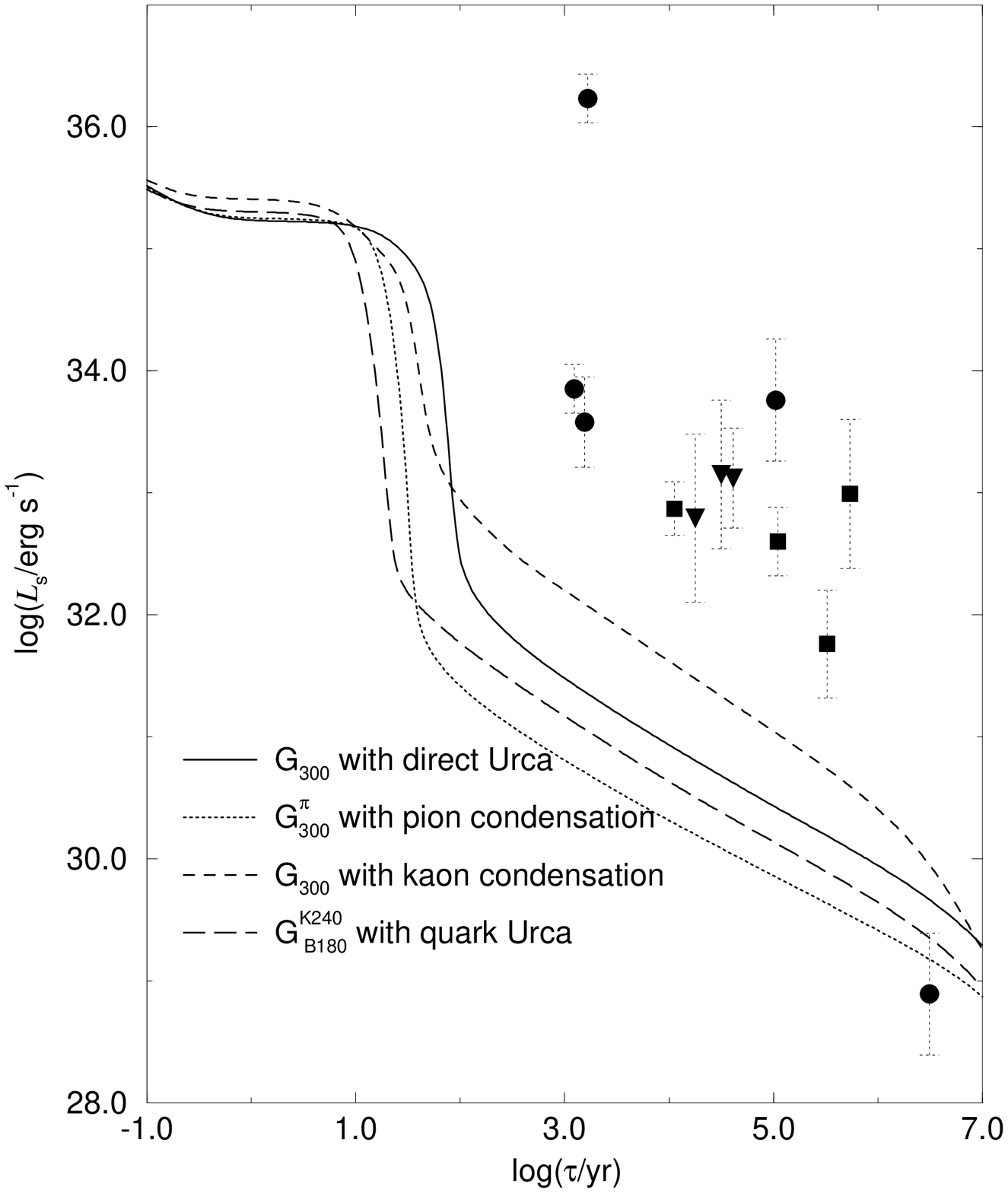,width=7cm}
  \capt{Influence of several different enhanced neutrino-emission processes
  on the cooling of neutron stars with masses $M=1.4\,M_\odot$, except
  for the kaon-condensed model where the mass is $M=1.78\,M_\odot$ (The
  $1.4\,M_\odot$ models do not possess large enough central densities
  to overcome the threshold density for kaon condensation).  The
  observed data are labeled in fig. \protect{\ref{fig:standard}}.}
  \label{fig:enhanced.nosf}
\end{figure}

Figure \ref{fig:enhanced.nosf} compares the influence of different
enhanced neutrino emission processes, i.e., direct Urca, pion and
kanon condensation, and quark Urca, on the cooling behavior of a
neutron star with mass $M=1.4\,M_\odot$. The nucleons and quarks are
treated as non-superflid particles. One sees that the inclusion of any
of these enhanced cooling processes reduces the star temperature too
quickly in order to get agreement with the observed data points.  The
only exception is PSR 1929+10. The most significant data points
(squares), however, lie considerably above the enhanced cooling
curves.
 
\begin{figure}[h] \centering 
  \psfig{figure=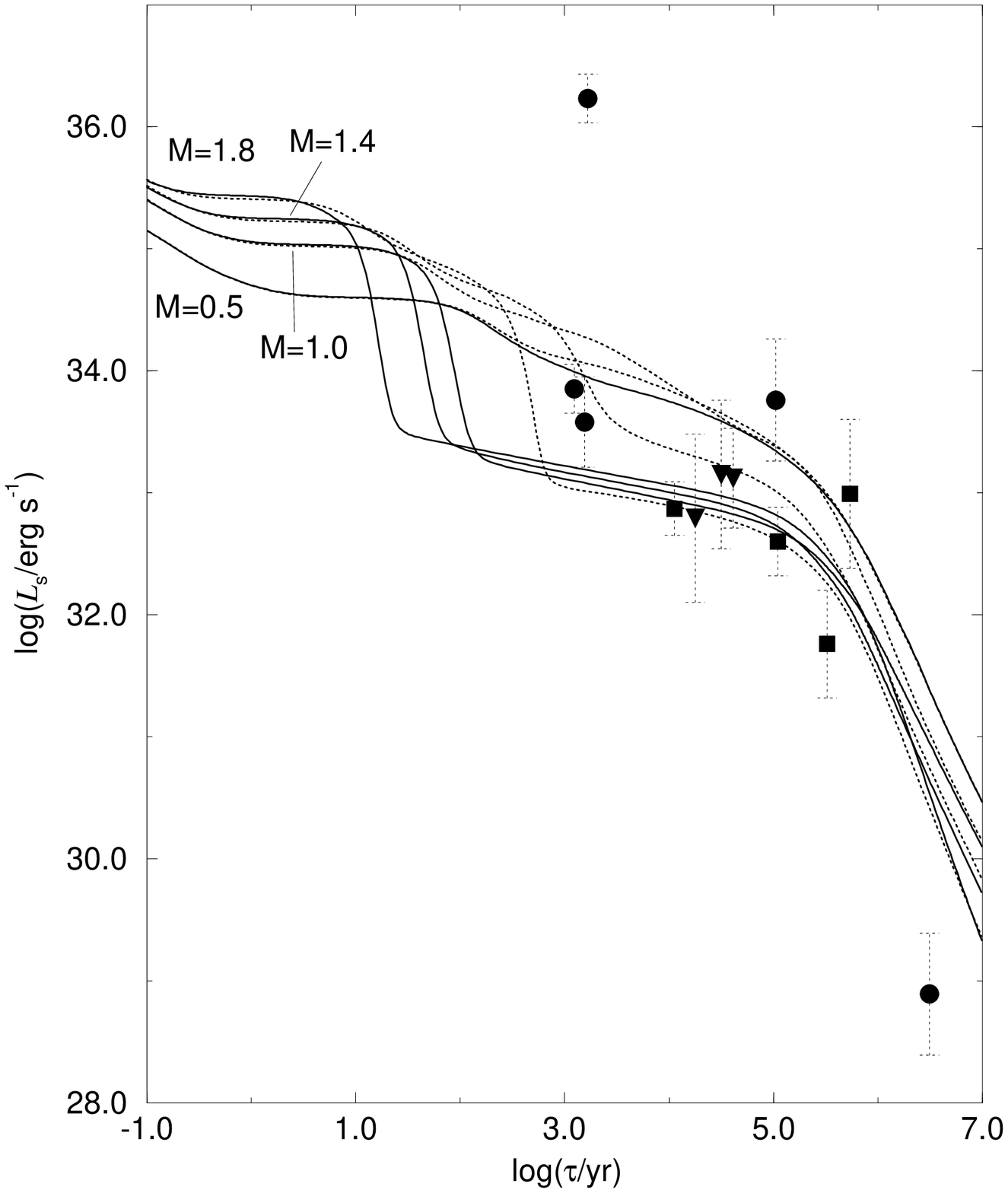,width=7cm}
  \capt{Influence of changes in the superfluid \sfp~ gap on the cooling of
  neutron stars constructed for \glen.  The enhanced cooling
  processes are direct Urca in pion (solid curves) and kaon
  condensed (dotted curves) matter.  The observed data are labeled
  in fig. \protect{\ref{fig:standard}}.}
  \label{fig:redsf}
\end{figure}

The enhanced cooling scenarios can be slowed down if one assumes that
the neutrons in the cores of neutron stars are superfluid, as was the
case in figs. \ref{fig:dr.uvu}--\ref{fig:k240b180}. The only problem
left then concerns the horizontal ``plateaus'' in these figures at
star ages between $10^2$ and $10^5$ years which tend to lie somewhat
too high for stars with canonical mass, $M=1.4\,M_\odot$, to be
consistent with the data. Note, however, that one gets agreement with
some of the data points if the star mass is assumed to be different
from the canonical value.  Another possibility to get agreement
consists in varying the gap energy, $\Delta_\mathrm{sf}$, which, as
mentioned in section \ref{sec:superfl}, is not very accurately
know. This is particularly the case for the high-density \sfp~
superfluid. A reduction of this gap by a factor of two, for instance,
moves the cooling curves up right into the region where the oberved
data is concentrated, as can be seen in fig. \ref{fig:redsf}.

Of course there are other possibilities, besides reducing the gap
energy, by means of which agreement with the observed data may be
achieved. For example, the superfluid phase of neutrons might not
reach the center of very massive neutron stars, as is the case for
stars constructed for \UVU~ (see fig. \ref{fig:dr.uvu}). This is due
to the smaller proton fraction and the resulting higher Fermi momenta
of neutrons.  Another possibility is an additional cooling process not
suppressed by superfluidity, as, for example, the superfluid
pair-breaking process \cite{Voskresenskii86,Schaab95b}, or 
so-called internal heating of neutron stars
\cite{Shibazaki89:a,Umeda93:a,Sedrakian93:a,Reisenegger95:a,VanRiper95:a}. 

\section{Summary}\label{sec:summary}

We have applied a broad collection of modern, field theoretical models
for the equation of state of superdense matter to the investigation of
the cooling behavior of compact stars, which comprise neutron stars
and their strange counterparts, the hypothetical strange stars.  The
models for the equation of state were derived elsewhere, using
many-body approximation techniques at different levels of
sophistication and complexity. Furthermore, uncertainties in the
behavior of matter at extreme densities, like the phenomenon of meson
condensation, superfluidity, hyperon degrees of freedom, transition to
quark matter, and absolute stability of strange quark matter are
incorporated in this collection.

Besides \emph{standard} cooling, which, subject to uncertainties in
the neutrino emission-rate of the modified Urca process, superfluid
gaps etc. (cf. section \ref{sec:con}), gives agreement between the
theoretical cooling curves and the observed data for some but not all
pulsars (the obtained luminosities are considerably too high in some
cases), we studied various \emph{enhanced} cooling mechanisms
too. These are connected to higher neutrino-emission rates coming from
the direct Urca process, $\pi$-- and $K$--meson condensates, and up,
down and strange quarks in the cores of neutron stars. It turns out
that the enhanced processes altogether cool the star models too
rapidly in order to get agreement with the data. So one is left with
introducing processes which delay somewhat the cooling as driven by
the enhanced processes. A plausible candidate is superfluidity, of
which there occur probably two different types in the cores of neutron
stars, namely \sfs~ and \sfp.  Due to the complicated theoretical
analysis, the density dependences and widths of the associated
superfluid gaps cannot be determined exactly, especially in the latter
case.  By using the more reliable \sfs~ gaps we find that already
small changes of the theoretical value for the \sfp~ gap shift the
cooling curves into the region of observed data, giving an overall
good agreement.  An accurate determination of the superfluid gaps in
neutron star matter seems therefore of particular interest.

Most interestingly, we find that--for a given star mass--not a single
model for the equation of state of our collection is able to explain
more than just a few data points!  This makes one wonder if the
observed data are possibly associated with stars of
\emph{different} masses. We recall that the canonical neutron star
mass is about $1.4\,M_\odot$, and the observed data
\cite{Thorsett93a} seem to concentrate about this value.
Nevertheless, the existence of pulsars with quite different masses,
e.g., neutron stars in binary systems that accrete mass from their
companions, seems equally plausible. If one assumes that the observed
luminosities are indeed coming from stars with different masses, those
of Vela and Crab could amount $\sim 1\,M_\odot$ and $\sim
1.4\,M_\odot$, respectively, then one may have a reason for the different
cooling behaviors.

As mentioned repeatedly, the absolute stability of strange quark
matter would have implications of greatest importance for various
branches of physics
\cite{Aarhus91}. It is agreed on that compact stars form ideal
environments where such matter could be found. Adopting a simple model
for the EOS of strange matter, we studied the cooling history of stars
made up of such matter, too. If the emissivities and heat capacities of
quark matter used here constitute reliable estimates, which, at
present, is an unresolved issue, we find that pulsars like Geminga,
Monogem, and PSR 1055-52 could be ruled out as strange pulsar
candidates. There is however one pulsar, PSR 1929+10, which could be
interpreted as a strange star. Of course, the QCD related
uncertainties in the properties of strange matter do not allow to draw
any stringent conclusions yet.  In order to provide a feeling for how
sensitive the cooling history of strange stars depends on the transport
properties of strange matter, we also performed cooling simulations
where the emissivity and heat capacity were slightly varied, due to a
possible superfluid behavior of the quarks.  As it turns out, this moves
the cooling curves right into the region of the luminosity--age plane
where most of the observed data are concentrated. So there is clearly a big
need for reliably determined transport properties of strange quark
matter.  At the present level of development it seems to us that no
definitive decision can be made as to whether the observed luminosities
are being emitted from conventional neutron stars or their strange
counterparts.

In summary, we find that really stringent conclusions about the
behavior of superdense matter on the basis of the observed cooling
data cannot be made at present. However this situation would change if
besides the luminosity and the age also the mass of just a single
pulsar could be determined! The rapid discovery pace of newly
observed pulsars in combination with the enormous progress that has
been made in observational X-ray astronomy are very encouraging in
this respect.

\vskip 2.truecm
\goodbreak
{\bf Acknowledgement:} \\

We would like to thank K. A. Van Riper for sending us his tables of
the mantle temperature as a function of luminosity, $T_m(L_m)$, which
were used in our cooling calculations.  Furthermore we have benefited
from conversations with Dany Page and Armen Sedrakian. This work was
supported by the Director, Office of Energy Research, Office of High
Energy and Nuclear Physics, Division of Nuclear Physics, of the U.S.
Department of Energy under Contract DE-AC03-76SF00098.

\clearpage


\clearpage

\end{document}